\documentclass[sigconf]{acmart}

\AtBeginDocument{%
  \providecommand\BibTeX{{%
    \normalfont B\kern-0.5em{\scshape i\kern-0.25em b}\kern-0.8em\TeX}}}

\copyrightyear{2025}
\acmYear{2025}
\setcopyright{cc}
\setcctype{by}
\acmConference[CHI '25]{CHI Conference on Human Factors in Computing Systems}{April 26-May 1, 2025}{Yokohama, Japan}
\acmBooktitle{CHI Conference on Human Factors in Computing Systems (CHI '25), April 26-May 1, 2025, Yokohama, Japan}\acmDOI{10.1145/3706598.3714057}
\acmISBN{979-8-4007-1394-1/25/04}

\def\projname{\textsc{IdeaSynth\xspace}}

\newcommand{\revise}[2][black]{\textcolor{#1}{#2}}

\definecolor{lightgrey}{rgb}{0.8,0.8,0.8}
\definecolor{white}{rgb}{1.0,1.0,1.0}

\usepackage{makecell}
\usepackage{longtable}
\usepackage{listings}
\usepackage{xcolor}
\usepackage{hyperref}
\usepackage{hyperxmp}

\usepackage{array}
\usepackage{amsmath}
\usepackage{microtype}
\usepackage{graphicx}
\usepackage{float}
\usepackage{wrapfig}
\usepackage{xspace}
\usepackage{enumitem}
\setlistdepth{9}

\usepackage{lipsum}
\usepackage{todonotes}
\usepackage{tikz}

\begin{document}

\title[\projname{} for Iterative Research Idea Development]{Iterative Research Idea Development Through Evolving and Composing Idea Facets with Literature-Grounded Feedback}


\author{Kevin Pu}
\email{jpu@dgp.toronto.edu}
\affiliation{
    \institution{University of Toronto}
    \country{Canada}
}

\author{K. J. Kevin Feng}
\email{kjfeng@uw.edu}
\affiliation{
    \institution{University of Washington}
    \country{USA}
}

\author{Tovi Grossman}
\email{tovi@dgp.toronto.edu}
\affiliation{
    \institution{University of Toronto}
    \country{Canada}
}

\author{Tom Hope}
\email{tomh@allenai.org}
\affiliation{
    \institution{Allen Institute for AI}
    \country{USA}
}

\author{Bhavana Dalvi Mishra}
\email{bhavanad@allenai.org}
\affiliation{
    \institution{Allen Institute for AI}
    \country{USA}
}

\author{Matt Latzke}
\email{mattl@allenai.org}
\affiliation{
    \institution{Allen Institute for AI}
    \country{USA}
}

\author{Jonathan Bragg}
\email{jbragg@allenai.org}
\affiliation{
    \institution{Allen Institute for AI}
    \country{USA}
}

\author{Joseph Chee Chang}
\email{josephc@allenai.org}
\affiliation{
    \institution{Allen Institute for AI}
    \country{USA}
}

\author{Pao Siangliulue}
\email{paos@allenai.org}
\affiliation{
    \institution{Allen Institute for AI}
    \country{USA}
}

\renewcommand{\shortauthors}{Pu et al.}
\begin{abstract} 
    Research ideation involves broad exploring and deep refining ideas. Both require deep engagement with literature.
    Existing tools focus primarily on broad idea generation, yet offer little support for iterative specification, refinement, and evaluation needed to further develop initial ideas.
    To bridge this gap, we introduce \projname{}, a research idea development system that uses LLMs to provide literature-grounded feedback for articulating research problems, solutions, evaluations, and contributions. 
    \projname{} represents these \emph{idea facets} as nodes on a canvas, and allow 
    researchers to iteratively refine them by creating and exploring variations and combinations.
    Our lab study ($N=20$) showed that participants, while using \projname{}, explored more alternative ideas and expanded initial ideas with more details compared to a strong LLM-based baseline.
    Our deployment study ($N=7$) demonstrated that participants effectively used \projname{} for real-world research projects at various ideation stages from developing initial ideas to revising framings of mature manuscripts, highlighting the possibilities to adopt \projname{} in researcher's workflows.
\end{abstract}


\begin{teaserfigure}
  \centering
  \includegraphics[width=0.8\textwidth]{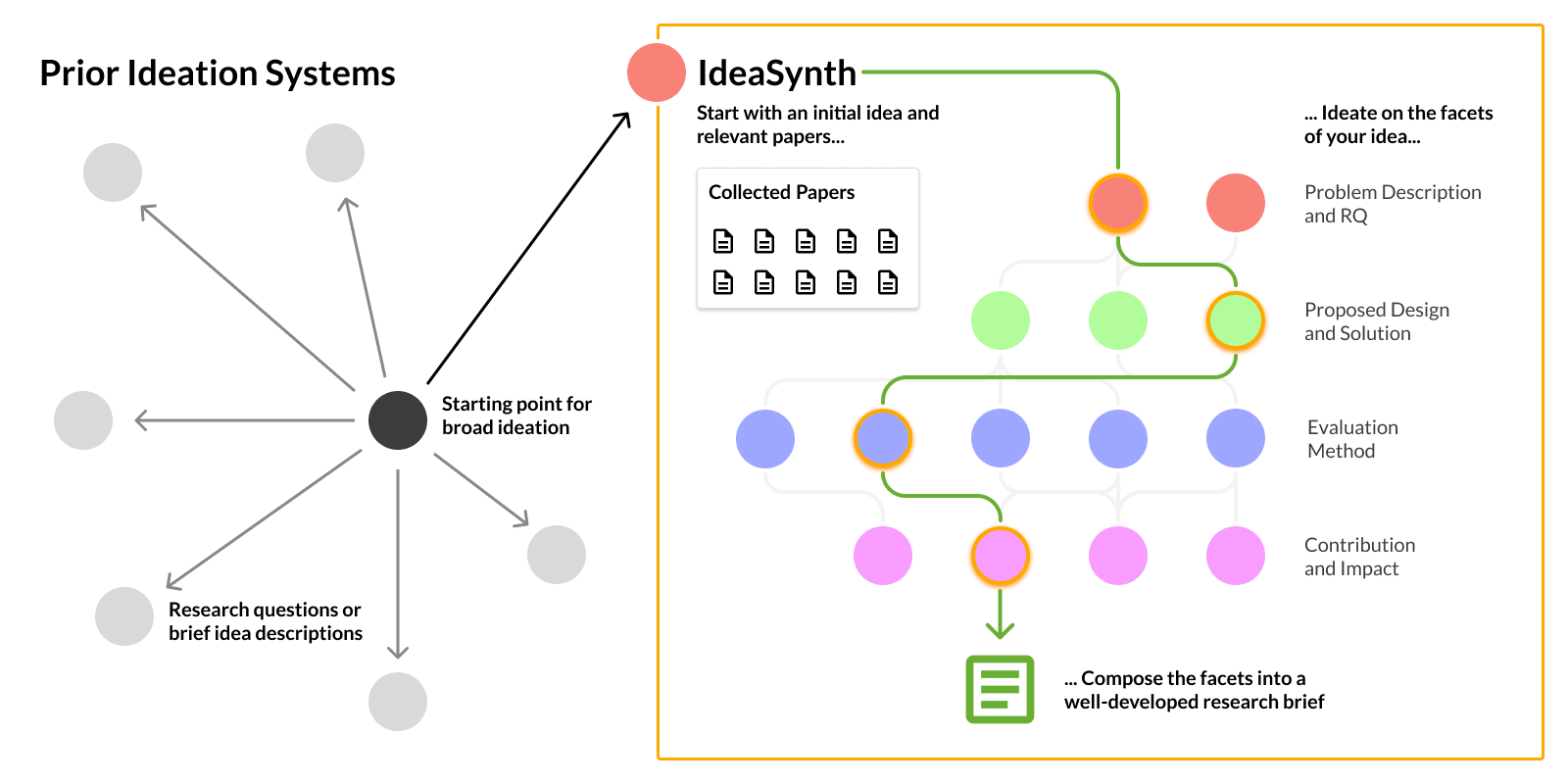}
  \caption{
  While prior systems focused on helping users explore a set of diverse ideas broadly, we present \projname{}, an ideation tool that helps users more deeply develop an initial research idea into a concrete and specific research brief. Intuitively, researchers often dissect projects into different facets to further develop them; \projname{} allows users to structure their idea as faceted idea nodes, explore and evolve different variations, and, finally, compose them into a complete research brief.
  }
  \label{fig:teaser}
\end{teaserfigure}

\maketitle

\section{Introduction}

Research ideation often starts with an initial spark of an idea \cite{Bell1993DoingYR}. However, the subsequent process of iteratively developing initial ideas, often high-level and under-specified, into a well-specified and structured research idea grounded by literature is equally important to the success of a research project \cite{vandenbroucke2018ideas, Ditta2016initialIdea, inie_how_2022, Sudheesh2016HowTW}. Unlike initial research ideation, this process is often \emph{multi-faceted} with researchers explore and consider many variations of research questions, methods, evaluation criteria, and contributions framing. Through this process, researchers iteratively converge on a specific, well-explored, high-impact, and feasible research idea \cite{vandenbroucke2018ideas,inie_how_2022}. However, this process is effortful, because it often involves both understanding of literature to generate variations for different facets \cite{lim2022advancing, Knopf_2006_doing_lit_review}, and also to \emph{compose} different idea facets together to form a coherent research idea \cite{Bell1993DoingYR}.

Recent work has shown great promise of leveraging Large Language Models (LLMs) to improve the initial stages of broad research ideation \cite{liu_coquest_2024, radensky_mixed-initiative_2024, Nigam2024AcceleronAT, Si2024CanLG, kang_synergi_2023,radensky2024scideator}. For example, tools that support literature sensemaking and discovery \cite{kang_synergi_2023}, generating new research questions and ideas \cite{liu_coquest_2024, Si2024CanLG}, or critiquing and evaluating ideas \cite{shaer_ai-augmented_2024, Si2024CanLG}. 
However, \revise{while prior works highlight LLMs’ usefulness in inspiring broad directions during early ideation, they} do not account for the \textit{further iteration and deeper development} phase of research ideation.
As a result, \revise{using LLMs to support} the process of focused expansion and distilling initial research ideas into a concrete and specific research idea is left largely under-explored. 

In this work, we investigate the user needs and design opportunities to support researchers in the \textit{expansion and refinement stages} of research ideation. Through a formative study with eight researchers, we identified three common pain points in researchers' workflows. First, researchers struggled with expanding initial ideas to concretely operationalize them into an executable project, which also hindered their ability to evaluate their ideas' novelty and feasibility. Second, researchers felt unsupported in organizing and evaluating multiple versions and iterations of their ideas. Third, researchers who have tried to use LLMs for research ideation often did not find the feedback helpful because it lacked the level of depth and specificity needed for refining and iterating on ideas. 

We address these three challenges by introducing \projname{}, a research idea expansion and refinement system that facilitates \revise{the user's process of} developing and experimenting with different variations of a research idea. 
\revise{\projname{} focuses on further evolving initial ideas by complementing human ideation with literature-grounded suggestions, guiding researchers to expand, explore, and connect their ideas, thereby enhancing the ideation workflow through human-AI collaboration.}
The system features a canvas interface that allows users to externalize variations of different \emph{idea facets}---such as research problem descriptions, proposed methods, and evaluation approaches---as nodes on the canvas. 
\projname{}'s LLM-powered idea refinement leverages the context of a user-controlled scientific literature collection to help the user refine individual facets and strengthen connections between facets.
Users can also create different \emph{idea paths} that connect different facets to generate a \emph{research brief}\footnote{In this paper, we refer to a research brief as a document that describes the necessary facets of a research project, before any project execution. We further define the term in Section \ref{s:scientific-ideation-process}.} 

To evaluate \projname{}, we conducted a controlled lab study and a field deployment study where participants used \projname{} to develop research ideas. In the lab study ($N=20$), 12 HCI and 8 NLP researchers developed ideas using \projname{} and a strong baseline system that shared a subset of \projname{}'s features. In addition, seven of the participants also completed a subsequent field deployment study where they used \projname{} for three days to develop research ideas from their own ongoing research projects. 
We found that \projname{} allowed participants to expand and iterate on their initial ideas, with details from the literature-grounded suggestions. Participants also felt that by externalizing idea facets and their connections, \projname{} supported them in considering more potential idea directions, avoiding idea fixation, and feeling more confident about their final idea. 
Through log analysis, we also identified behavioral differences where participants in the baseline condition spent more time on writing based on their prior knowledge and perceived the baseline AI assistance lacking for supporting ideation.
Further, participants in the field deployment study used \projname{} across various stages of ideation, from initial project specification to framing exploration on manuscripts. They saw it fitting into their existing workflows by lowering the barrier for ideation, and also envisioned using it collaboratively with others. 
We highlight the implications of our findings by discussing the need to tailor AI support for different research processes and diverse user needs.
\revise{We also identify potential limitations and ethical concerns of involving LLM in the research ideation process, presenting caveats for future system designs.}

In summary, this paper makes the following contributions:
\begin{itemize}
    \item   A formative study that illustrates challenges researchers face during ideation, including exploring different variations of ideas, expanding on their initial insights, and eliciting feedback from LLMs that help them to evaluate and refine their ideas.
    \item  Based on the above, we introduce \projname, a research ideation tool that helps users expand and develop their initial ideas, with:
  \begin{enumerate}
    \item \revise{actionable, literature-grounded LLM-powered feedback for idea expansion, refinement, and evaluation
    }
    \item \revise{a canvas editor that allowed users to structure their research idea as facet nodes and explore different variations, alternatives, and compositions.}
  \end{enumerate}
    \item   A lab and a field deployment user study demonstrating how \projname{} can assist users in expanding and refining research ideas and how its designs and interactions can \revise{enhance} real-world research ideation workflows.
\end{itemize}

\section{Background and Related Work}

\subsection{Scientific Research Ideation process}
\label{s:scientific-ideation-process}
Scientific research ideation is a non-linear complex process consisting of multiple scientific tasks and activities~\cite{Foster2005nonlinear_info_seeking}. To come up with a new idea, researchers have to identify and prioritize interesting research problems, form ideas to address selected problems, compare the idea with existing literature, integrate new knowledge to make sense of the relevant information space, and develop an evaluation plan~\cite{hope2023computational}. These tasks appear in different phases of a research project and involve different cognitive processes.

In the beginning phase of scientific research ideation, researchers mostly engage in problem identification and prioritization. This early phase can involve ``opening processes'' of information-seeking which includes both intentional and serendipitous starting points~\cite{Foster2005nonlinear_info_seeking}. Researchers generate and evaluate initial research questions~\cite{hope2023computational}. Like in a typical ideation process, the initial phase is associated with the divergent stage~\cite{paulus2000idea} where researchers can benefit from exploring various research questions before committing to a selected few~\cite{liu2023creative,liu_coquest_2024}. The outcome of this phase is mostly a set of prioritized research questions that will be further explored.

After researchers decide on an idea to pursue, they start shaping the idea into a research brief, sometimes called a research proposal. In this paper, we define research brief as a description of a research project before any execution (e.g., running an experiment, implementing a system).
A research brief helps make the research project more concrete and is often used to communicate the ideas to others for feedback. 
A research brief can include multiple ideas, ranging from a research question or problem, a method, a hypothesis or antithesis, and a theory~\cite{inie_how_2022}. For example, a research brief can have a problem, a method, and a hypothesis with an experimental design.
This research-brief-forming phase involves "orientation processes" where researchers make their ideas more grounded and concrete with literature-based evidences and details~\cite{Foster2005nonlinear_info_seeking,hope2023computational}. Researchers often consult existing literature to situate their ideas, make the problem more specific, generate idea facets (e.g., method, evaluation criteria), and evaluate idea facets and the resulting ideas \cite{lim2022advancing, Knopf_2006_doing_lit_review}. This phase of idea generation, typically associated with the converge stage~\cite{paulus2000idea}, is especially important for research ideation because of the potentially high costs of a research project. A well-thought of research brief can ensure that resources are spent on high-impact problems and facilitate idea communication.

\subsection{Research ideation support with AI}

\subsubsection{Initial Idea Generation}
Recent work on ideation informs us of the potential and limitations of leveraging AIs for ideation support. For example, prior work suggests that LLMs or humans with access to LLMs can generate more novel ideas~\cite{Si2024CanLG,qi2023large,yang2023large} and higher quality ideas~\cite{girotra2023ideas} than humans who do not use LLMs. While the majority of prior work in this space suggests that providing AI support during ideation is beneficial to users~\cite{he_ai_2024,qi2023large,girotra2023ideas,liu_coquest_2024,hope2023computational}, a few also pointed to potential risks of incorporating AI in ideation, such as generating less original ideas~\cite{wadinambiarachchi2024effects}; less diverse ideas~\cite{anderson2024homogenization,wadinambiarachchi2024effects}; and more fixation~\cite{wadinambiarachchi2024effects}. In sum, prior work points to how carefully design UIs and interactions may be keys to successful integration of AIs into the ideation process, and that failure to do so may incur severe adverse effects.



More directly related to our work, multiple AI systems and interventions have been tailored to support the initial broad ideation of research ideas where ``opening processes'' happen. CoQuest helps researchers generate research questions with an LLM-based agent~\cite{liu_coquest_2024}. 
Many prior systems provided users with relevant inspirations instead of generating the ideas directly. For example, helping users find relevant research analogies as inspirations~\cite{kang_augmenting_2022}; generating personalized inspirations using LLMs with an evolving knowledge graph~\cite{gu_generation_2024}.
However, most of prior systems only focused on supporting initial broad ideation instead of further developing them into a concrete research brief. For example, CoQuest only supports generating research questions, but does not consider potential solutions nor how to evaluate them to demonstrate impact and contribution.


\subsubsection{Idea Convergence, Evaluation, and Refinement}

Some systems focus support in the converging stage for evaluating research ideas.
Prior work has proposed ways to provide users with feedback for generated ideas.
Shaer et al. explores the application of LLMs to the converging phase of brainwriting by using LLMs to help evaluate the ideas based on their novelty, relevance, and insightfulness~\cite{shaer_ai-augmented_2024}. 
Liang et al. and D'Arcy et al. separatedly shows that language models can generate helpful feedback on scientific papers~\cite{liang2023can,darcy_marg_2024}. 
Lu et al. reports that automated paper evaluation has higher inter-rater agreement than human evaluation~\cite{lu2024ai}. 
Shen et al. provides AI explanation dialogues to support human-AI scientific writing \cite{shen_convxai_2023}.
These findings suggest the potential for automatically evaluating ideas. However, there are contradicting findings that caution against relying too heavily on automated idea assessment. 
LLMs still cannot evaluate research ideas consistently~\cite{Si2024CanLG}. Moreover, evaluating research ideas is a highly subjective task that can be affected by evaluators' preferences and experience. Human experts do not always agree on the novelty of a research idea~\cite{Si2024CanLG}. Higher inter-rater agreements of LLM ratings~\cite{lu2024ai} do not mean that the generated ratings capture the intrinsic quality of an idea. 
Moreover, the format of research ideas that get evaluated in prior work are mostly short text description~\cite{liu_coquest_2024} or a writeup on the formal level of a published paper ~\cite{chamoun2024automated, liang2023can}.
Informed by these prior work, instead of producing evaluation scores, \projname{} gives actionable suggestions that help users decide on what to do to improve their ideas. In contrast to full research papers describing a completion work, we aim to help user produce research briefs that are less formal and more succinct compared to full papers. Yet, compared to broad research ideas that most prior work aimed to target (e.g., generate research questions), research briefs are more specific and detailed, which have different implications for evaluation.

Another research thread explores how to help researchers refine existing ideas. SCIMON compares a given research idea, specified by faceted context (e.g., methods, problems), to existing literature and automatically updates the idea if it is too similar to existing work~\cite{wang_scimon_2024}. ResearchAgent, an LLM-based research idea writing agent, generates research ideas by adding an idea facet at a time starting from problem, method, to experiment design~\cite{baek_researchagent_2024}. ResearchAgent also provides review, feedback and ratings of an idea and automatically updates some ideas based on the feedback ~\cite{baek_researchagent_2024}. While ResearchAgent divides idea generation into facets, they have to be generated in a fixed sequence, unlike in ~\projname{} that gives users the freedom to generate idea facets in any order. ~\projname{} also allows users to apply context to specific facets instead of the whole idea. Acceleron, an LLM-based system provides a prompt template that helps users rewrite a research proposal based on identified gaps in literature~\cite{Nigam2024AcceleronAT}. Similarly, PaperRobot \cite{wang_paperrobot_2019} incrementally generate paper titles and abstracts based on existing papers. 
However, Acceleron and PaperRobot generate research artifacts with an initial idea description and little human involvement, which is different from the approach of supporting researchers develop a more detailed research brief in \projname{}.

\section{Formative Study}
We conducted a formative study with 8 HCI and NLP researchers (5M, 3F) both to understand the challenges in existing ideation workflows and also to identify design opportunities for enhancing research ideation with AI.
The study consists of a walk-through of participant's existing artifact of a past or ongoing research project (e.g., research project proposals and running project notes documents), followed by a live ideation session using a Wizard-of-Oz (WoZ) prototype that leveraged an off-the-shelf research support tool. Finally, we conclude the study with a semi-structured post-interview.

We recruited Ph.D. students and postdoctoral researchers through our organization's Slack channels and through personal connections.
We focused our recruitment on researchers who were actively involved in project ideation and could provide detailed accounts of their workflows.
Demographic details of our participants can be found in Appendix \ref{a:formative-participants}. We selected participants on a first-come, first-serve basis, and kept recruitment open during our study until we reached data saturation.

All studies were facilitated by two members of the research team and conducted virtually via Google Meet. Each study was around 60 minutes in length and was divided into 3 parts. To ground the study to real-world research projects, prior to the study, we asked participants to submit a project document---a running artifact with project updates, ideas, links, meeting notes, and more. This artifact was the central focus of \textbf{Part 1} of the study, where we asked participants about how they sourced and developed their ideas, challenges with ideation, and their use of AI tools in research.

In \textbf{Part 2}, participants completed an interactive activity with a WoZ design probe. 
We leveraged Google Docs\footnote{\url{https://docs.google.com/}} and Perplexity AI\footnote{Perplexity AI with Academic Focus has capability of conducting web search focused on scientific publications: \url{https://www.perplexity.ai/}} to power the ``system interface'' and ``LLM-based intelligent support'' aspects of our WoZ study, respectively.
Specifically, Google Docs' real-time collaboration features enabled the study facilitator to populate content to the shared document in real-time from a different computer; and
Perplexity AI's ability to search the web and provide functional links for academic papers could help us better understand the design opportunities of a research ideation system connected to an academic database.
The probe's interface consisted of two side-by-side Google Docs. One of the Docs (the ``ideation doc'') contained three initial research ideas we pre-generated by prompting Perplexity AI with participants' research interest from the recruitment survey as well as attachments of 1--3 of their recent publications (\S\ref{a:formative-prompt}). The other Doc (the ``output doc'') is used as a WoZ AI-ideation support system that displayed AI outputs.
During the study, participants could trigger an ``AI assistant'' in the ideation doc by prefacing a request with a ``!'' command, after which a study facilitator (i.e., one of the authors) would manually forward the request to Perplexity AI and then paste its output back into the output doc. Participants spent around 25 minutes working with the probe to further develop a research idea. They had the freedom to build on one of the three ideas presented to them, use some combination of those ideas, or ignore them altogether. This activity ended when the participant wrote a short paragraph about their idea outlining aspects that may include a research question, methods, related work, and expected outcome.

We concluded with \textbf{Part 3}, an exit interview in which participants reflected on their experience using the design probe and how it differed from their typical workflows. They also shared feedback on the probe and discussed desirable ways of interacting with AI in ideation support tools.

After the formative study, the first author analyzed the transcripts with open coding followed by thematic analysis \cite{Boyatzis1998TransformingQI,Connelly2013GroundedT} to identify key themes, focusing on common challenges participants faced. Tentative themes were discussed and iterated upon with the larger research team at weekly meetings to arrive at our final set of themes. The codebook can be found in Appendix \ref{a:codebook}. Below, we describe the common themes we found in our analysis:


\subsection{Challenge 1: Expanding Initial Ideas to Concrete Projects}
\label{s:formative-c1}
While prior works on ideation emphasized on supporting broad ideation and exploring diverse directions~\cite{liu_coquest_2024, shaer_ai-augmented_2024}, participants from our study also point to scenarios where they were already settled on a broad direction but struggle to further develop them into a concrete research project.
Further, many said that they often do not lack interesting broad initial research ideas to pursue. However, they pointed to how it is often more challenging to expand their initial ideas due to uncertainty over how to concretely operationalize them. Participants also highlighted difficulties in measuring the novelty of their ideas as they often have broad topics of interests but without specific idea facets (e.g., designs and methods, how to evaluate, specific impacts, and broader contributions) to solidify the project. P2 felt \textit{``initially there's a lack of clarity into what the proposed approach would look like so it's hard to talk about it in a very concrete way. Because I am not sure what the core components of my approach will be. It's hard to even express it in writing.''} 
Moreover, the process of narrowing down a broad range of thoughts to an operationalized project idea is inherently difficult. P3 expressed that \textit{``narrowing it [initial idea] down to the one operationalizable research hypothesis...is what takes the maximum amount of effort.''} 

Understanding existing literature deeply is crucial for evaluating ideas but poses significant challenges. As P2 noted, starting a project without a comprehensive literature review makes it hard to confidently discuss existing gaps. 
The overwhelming volume of related works make it even more difficult to ensure all relevant literature is covered (P4).
Researchers often feel insecure about the novelty and impact of their ideas, relying on validation from peers and advisors, who may point out limitations in impact or feasibility (P5).
This presents the need for literature-grounding and feedback to contextualize and iterate during the idea expansion process.

\subsection{Challenge 2: Exploring Variations and Iterations of Ideas}
When further developing an initial idea into a concrete research project brief, participants reported a common strategy of exploring variations and different framings. However, they faced challenges in evaluating multiple versions and iterations of their ideas. P6 often developed early prototypes of several ideas, setting aside those that do not work for later reconsideration.
They often present multiple variations to peers and advisors to get feedback, which allowed them to identify blind spots and pivot between directions. P5 described that they will \textit{``try to have more than one base idea, like two or three, just so that when I do get the opportunity to talk with my advisors...if one is a dead end we can talk about others as well...I try to create a few variations of each base idea.''}

This iterative process often involves extensive documentation, such as slide decks with updates and meeting notes, long-running documents with figures and summaries, and detailed research proposals with multiple versions and planning scenarios (P4, P5). Managing the context of multiple ideas stemming from the same seed idea is particularly challenging, as researchers struggle to track various developments and inspirations that are not immediately within the current scope (P1). The spatial fragmentation of notes and the need to link external resources further complicate the organization and evaluation of ideas (P1).

\subsection{Challenge 3: Receiving Useful Support from LLMs}
Reflecting on interacting with our WoZ prototype during the study and their own experience using LLM-based tools for ideation, participants felt that existing tools do not provide a useful level of specificity and literature-relevance in their feedback.
For example, P5 felt that the AI often reiterates the same points without adding substantial new insights, failing to clarify concepts or identify specific details needed to refine ideas. P4 proposed that researchers would benefit from AI providing fair and critical feedback, asking research questions, and offering critiques rather than simply affirming ideas.
Furthermore, using AI for literature review tends to lead to reiterations of known information without providing the motivational context or the ``why'' behind certain approaches, which is crucial for ideation (P3). 

Interaction paradigm with AI is another area of concern. While most participants ($n=5$) agreed that conversational interaction with AI is useful, typing natural language requests may not always be the preferred interaction method. Instead, participants desired alternative designs where AI support could be more active and in-situ within their artifacts and workspaces but without being disruptive (P2, P5). 
Additionally, AI suggestions often lack literature context or history, making it hard for researchers to see and understand the relevance of the suggestions (P7).
In terms of agency and ownership, participants highlighted that they prefer using AI to inspire ideas, assist in study design, and spark further development rather than generating final outputs (P3, P5).


\subsection{Design Goals}
\label{s:design-goals}
Based on the challenges identified by our formative study, we formalize three design goals corresponding to each of the challenges listed above:
\begin{itemize}
    \item \textbf{DG1:} Better support for further developing an early stage research idea into a concrete research brief through reflection and evaluation.
    (\textbf{Challenge 1}).
    We observed the need for support with idea \textit{expansion} rather than idea \textit{generation}. We thus aim to provide better tools to make ideas more detailed and concrete so that researchers can operationalize them.
    \item \textbf{DG2:} Help users explore and evaluate different variations of an idea by analyzing their strengths and weaknesses and visualizing the connections between ideas facets (\textbf{Challenge 2}). Researchers found it difficult to manage idea iterations and fragmented notes while working in linear artifacts (e.g., documents, slides). New, non-linear interfaces may be needed to help researchers visualize and make sense of relationships between various idea components. 
    \item \textbf{DG3:} Enhance the specificity and relevance of LLM support by grounding the response in relevant information in literature and focusing on inspiring users to develop their idea further (\textbf{Challenge 3}). Researchers often found LLM outputs too generic to be useful. This may be because LLMs did not encode the necessary domain-specific knowledge during training or require user guidance in accessing less well-represented knowledge in training data. We thus increase relevance and specificity by leveraging users' existing collections of related literature to ground LLM responses.
\end{itemize}
\section{\projname{}}

\begin{figure*}[h]
    \centering
    \includegraphics[width=\textwidth,keepaspectratio]{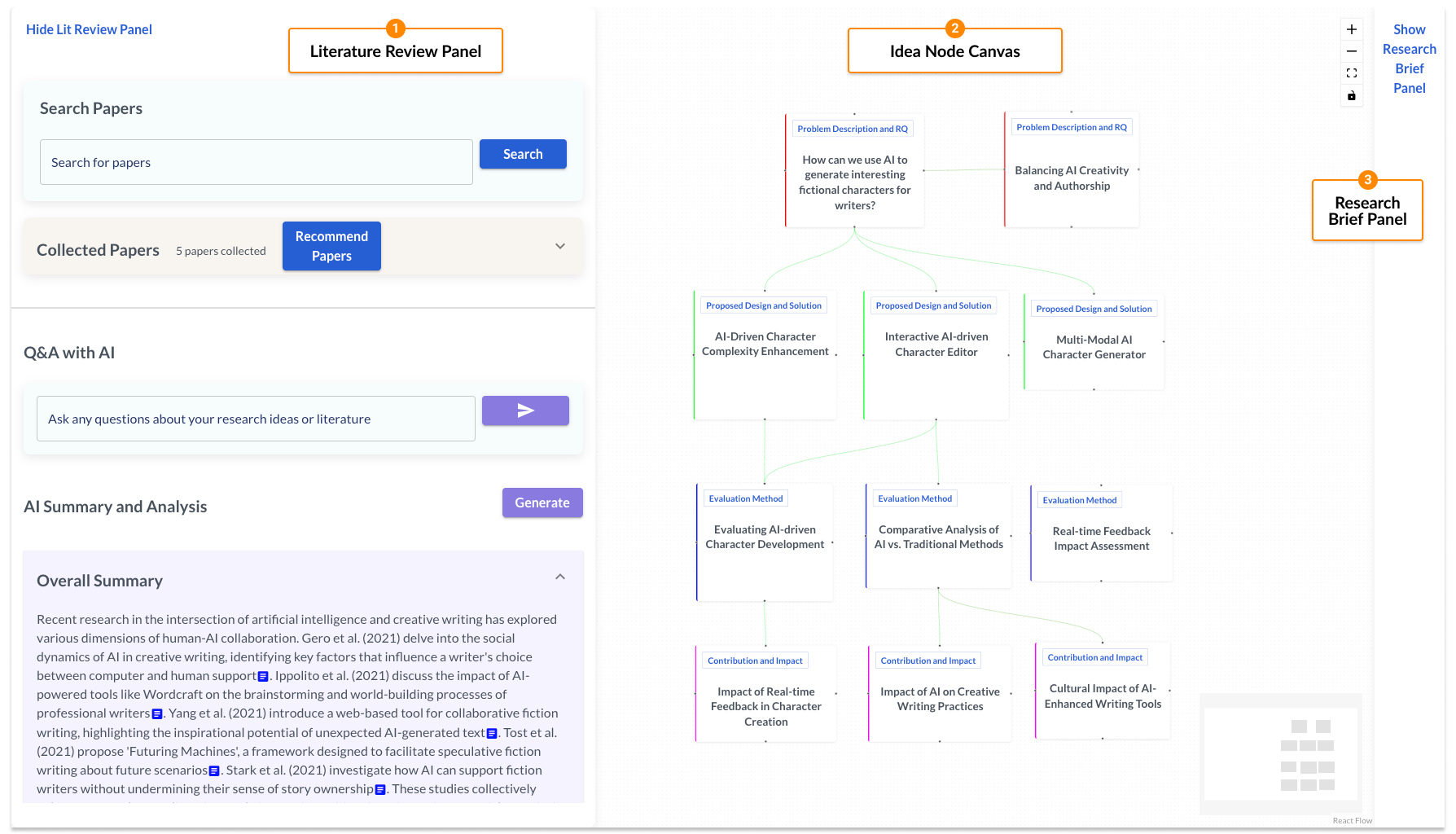}
    \caption{
        \textbf{\projname{}'s system UI screenshot.}
        \textnormal{On the left is the Literature Review Panel \textcircled{\raisebox{-0.5pt}{1}}, where the user can search and add papers to the collection, and generate literature summary and analysis. In the middle is the Idea Node Canvas \textcircled{\raisebox{-0.5pt}{2}}, where the user can externalize idea facets as nodes and explore and expand ideas in a tree-like hierarchical structure. The nodes are semantically zoomed out to display the idea canvas. Full UI and functionality of the node can be found in Figure \ref{fig:node-UI-menu}. On the right is the Research Brief Panel \textcircled{\raisebox{-0.5pt}{3}}, the user can select multiple nodes and generate a coherent research brief that describes the selected ideas facets (Expanded UI in Fig.\ref{fig:research-brief-panel}).}
        }
    \label{fig:UI-screenshot}
\end{figure*}

Based on our design goals, we present \projname{} (Fig. \ref{fig:UI-screenshot}), a research idea development system that helps scholars expand and refine their initial ideas. The system represents idea facets as linked nodes on a canvas, visualizing the logic relationships and forming a tree-like information structure for the user's ideation context (DG2). \projname{} also provides contextualized feedback at the node level and for the overall idea canvas based on the collected papers (DG3), enabling users to iteratively enrich ideas, explore alternatives, and connect facets to create a coherent research brief (DG1).
We chose the research brief as the system's output due to its utility in communicating and refining research projects, as it encapsulates core ideation activities such as defining a problem, method, and hypothesis.


\subsection{User Scenario and System Walk-through}
In this section, we use a user scenario to walk through various features and design of \projname{}.

\subsubsection{Literature-based broad ideation.}

An HCI researcher is brainstorming a new project on \emph{using AI to assist with fictional character development in creative writing}. She begins with a literature review, which helps refine her research question to: \textit{``How can we use AI to generate interesting fictional characters for writers?''}. However, she faces challenges in narrowing the project scope to propose novel contributions and worries about prematurely focusing on a potentially underdeveloped idea.

\begin{figure*}[!t]
    \centering
    \includegraphics[width=\textwidth, height=\textheight, keepaspectratio]{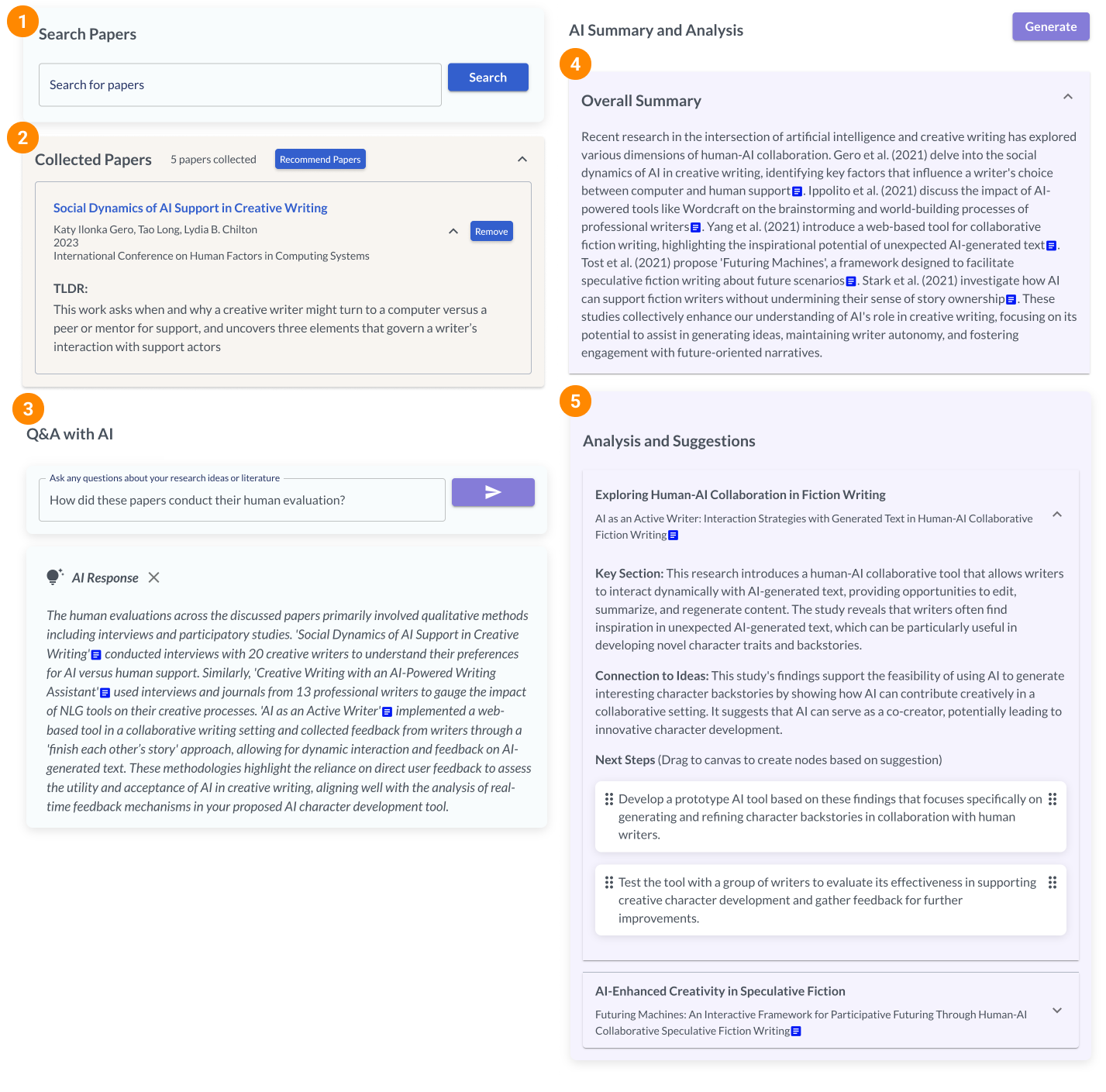}
    \caption{
        \textbf{Literature Review Panel.}
        \textnormal{This is a scrollable side-panel. The figure expands the panel to two columns (flow from left to right) to display all components. The user can \textcircled{\raisebox{-0.5pt}{1}} search and add papers, and use \texttt{Recommend Papers} to expand the collection \textcircled{\raisebox{-0.5pt}{2}}. They can \textcircled{\raisebox{-0.5pt}{3}} pose targeted questions about details in the paper or inquire about their idea nodes. On the right column, the user can generate an overall literature summary based on the paper collection \textcircled{\raisebox{-0.5pt}{4}}, and an analysis that connects key paper sections to the idea nodes on the canvas \textcircled{\raisebox{-0.5pt}{4}}. In our evaluation, the baseline system contains only the functionalities in the left column (Section \ref{s:evaluation-condition}).}
    }
    \label{fig:lit-review-panel}
\end{figure*}

\subsubsection{Decomposing initial idea into idea facet nodes and adding relevant papers.}
\begin{figure*}[!th]
    \centering
    \includegraphics[width=\textwidth,keepaspectratio]{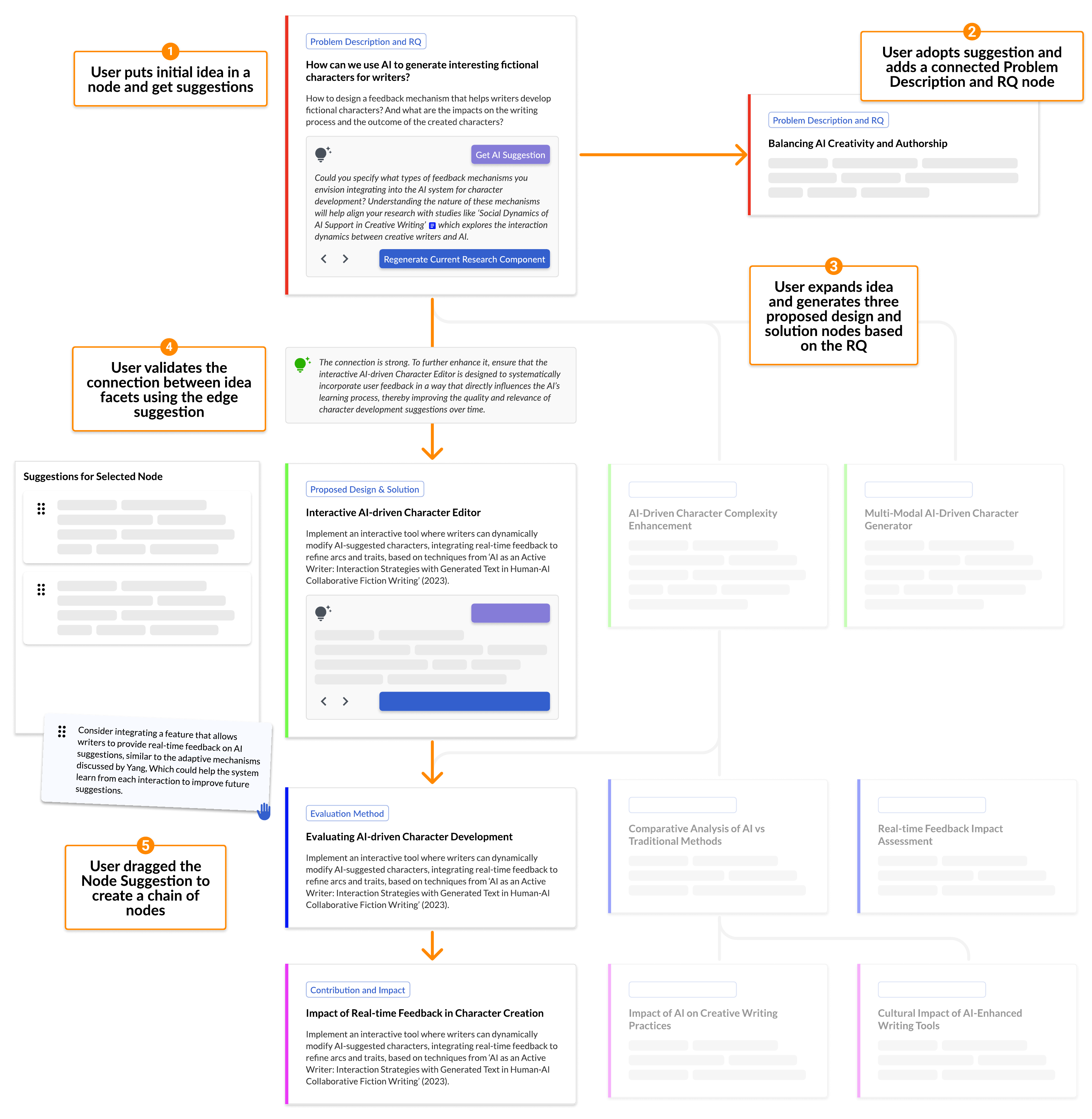}
    \caption{
        \textbf{Node Canvas UI demonstrating the user scenario workflow.}
        \textnormal{The HCI researcher starts by entering her initial idea in a blank node \textcircled{\raisebox{-0.5pt}{1}}. Then, she uses \texttt{Get AI Suggestions} to receive feedback. \projname{} prompts the researcher to clarify the problem description and scope, and also proposes an alternative research question. She adopts the suggestion and generate another node \textcircled{\raisebox{-0.5pt}{2}}. To further expand on her idea, the researcher explored other idea facets and generated three nodes on proposed design and solution \textcircled{\raisebox{-0.5pt}{3}}. She evaluated the nodes and used the edge to validate the connection between the design and the problem \textcircled{\raisebox{-0.5pt}{4}}. To examine the particular node, the researcher expands the left arrow menu to view node-based literature analysis that connects the idea facets to key sections from the collected papers. The node literature analysis also offers actionable suggestions, which she uses to drag and create a chain of subsequent nodes to ideate on the evaluation and contribution \textcircled{\raisebox{-0.4pt}{5}}.}}
    \label{fig:canvas-UI}
\end{figure*}

The researcher decides to use \projname{} to further develop and refine her research idea.
She first created a ``Problem Description and RQ'' node on the canvas interface and entered her research question (Fig.\ref{fig:canvas-UI}.\textcircled{\raisebox{-0.5pt}{1}}). This node represents a ``facet'' of the idea. Then, using the paper search function, she adds the papers she collected during her literature review into the paper collection tray (Fig.\ref{fig:lit-review-panel}.\textcircled{\raisebox{-0.5pt}{1}}\textcircled{\raisebox{-0.5pt}{2}}). This way, \projname{} could better understand her mental model and provide more contextualized feedback as she continue to develop her idea. She also uses the ``Recommend papers'' button to find additional relevant papers and expand her literature review (Fig.\ref{fig:lit-review-panel}.\textcircled{\raisebox{-0.5pt}{2}}).
She can add as many papers as she wants as long as each paper's content and generated summaries fit the LLM context window (around 300 pages for \texttt{gpt-4-turbo}).

To obtain an overview of the collected papers, she generates an AI literature summary and analysis (Fig.\ref{fig:lit-review-panel}.\textcircled{\raisebox{-0.5pt}{4}}\textcircled{\raisebox{-0.5pt}{5}}). The system provides a one-paragraph summary and a tailored analysis that draws connections between nodes on the canvas and specific sections of relevant papers. From the summary, she gains a deeper understanding of the challenges around evaluation from a prior work that also focused on AI writing assistants and explored mechanisms of using AI to drive fictional character development. Clicking on the paper icons in the summary opens the referenced papers in the collection tray for further reading. 
Beyond showing connections between prior work and the canvas, the literature analysis also provided actionable next step suggestions for expanding the ideas on the canvas. For example, she could drag a suggestion block onto the canvas to generate a chain of nodes that represent a new idea. 
For now, she decides to first explore and expand upon her initial research question.

\subsubsection{Further developing the idea by exploring facet node variations.}
Returning to her initial RQ facet node, she writes down two more detailed research questions in the node: \textit{``How to design a feedback mechanism that helps writers develop fictional characters?''} and \textit{``What are the impacts on the writing process and the outcome of the created characters?''} (Fig.\ref{fig:canvas-UI}.\textcircled{\raisebox{-0.5pt}{1}}). She uses \texttt{``Get AI Suggestions''} to receive feedback. \projname{} first presents a clarifying question to ask the researcher to specify what aspects of the writing process she would like to focus on, with examples such as \emph{initial character creation} or \emph{continuous character development throughout writing}. The suggestions also reference relevant papers with findings on how writers seek feedback.

Based on this suggestion, she iterates on her research question to center around \textit{``character creation aligned with a plot narrative vision in early writing''}. She also reads a suggested paper in-depth and gains insights on existing writer feedback mechanisms. She then toggles to view another AI suggestion. This time, \projname{} proposes an alternative research question to consider \emph{how to balance AI creativity and authorship}, a key issue illustrated by a paper she had previously added to the system. The researcher finds the research question inspiring as she has previously not considered the trade-off in authorship. This time, she uses \texttt{``Generate Alternatives''} feature to create new nodes that explore variations of this suggested research question. 
She incorporates the generated node on \emph{investigating the author's creative control and authorship} and connects it with her first node (Fig.\ref{fig:canvas-UI}.\textcircled{\raisebox{-0.5pt}{2}}).

\subsubsection{Creating connections between nodes to explore their relations.}
She then uses the downward arrow menu at the bottom of the two ``Problem Description and RQ'' nodes to explore other idea facets (i.e. ``Proposed Design and Solution'', ``Evaluation Method'', and ``Contribution and Impact''). First, she uses \projname{} to generate multiple ``Proposed Design and Solution'' nodes with the prompt \textit{``interactive AI system for character building with author feedbcak''} (Fig.\ref{fig:canvas-UI}.\textcircled{\raisebox{-0.5pt}{3}}). 
The generated nodes are color-coded by idea facet and linked to their original nodes. The researcher evaluates each design: she finds the first two compelling—one is a character editor with real-time AI feedback, and the other enhances character complexity by analyzing literary databases. She finds the third, a multi-modal character generator, interesting but beyond the project's scope and deletes it. She then examines the connecting edges between nodes to validate their relevance to the original research questions, checking the logical strength of the node connection and receiving suggestions for improving it (Fig.\ref{fig:canvas-UI}.\textcircled{\raisebox{-0.5pt}{4}}). The researcher sees that the connections to the two proposed designs are strong, and adopts one of the edge suggestions by further specifying how the AI learns and incorporates writer's feedback.

\begin{figure*}[h]
    \centering
    \includegraphics[width=\linewidth]{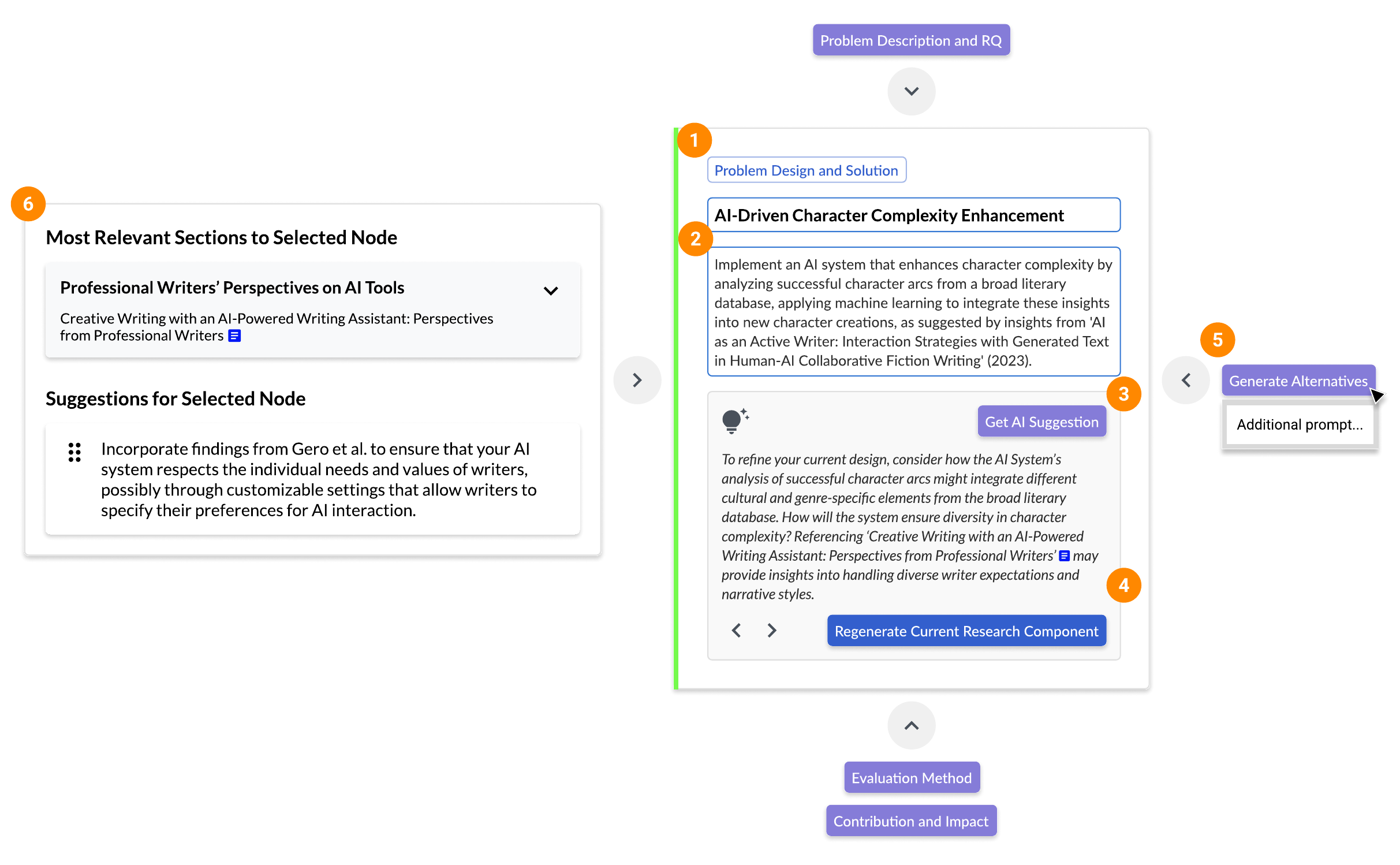}
    \caption{
        \textbf{Idea Node UI.}
        \textnormal{In \projname{}, idea facets are represented as nodes. The user can select the type of idea facet \textcircled{\raisebox{-0.5pt}{1}} using a drop-down menu, the left border is color-coded based on the facet type. The user can enter a node title to represent the key theme of this node, and describe the idea in more details in the node content text field \textcircled{\raisebox{-0.5pt}{2}}. The user can view node-level AI suggestions in the form of clarifying questions, idea alternative, and idea expansion \textcircled{\raisebox{-0.5pt}{3}}. They can manually adopt the suggestion to use the action button to let \projname{} execute \textcircled{\raisebox{-0.5pt}{4}}. On the four sides of the node, there are expandable menus. User can use the purple buttons to generate idea facet nodes in a hierarchical structure. The user can optionally enter a prompt to steer the generation \textcircled{\raisebox{-0.5pt}{5}}. The left expandable menu reveals the node-level literature analysis, which connects the idea facet to relevant paper sections and provide suggestions \textcircled{\raisebox{-0.5pt}{6}}. }
    }
    \label{fig:node-UI-menu}
\end{figure*}

The researcher can also use the left arrow menu to review an automatically generated literature analysis specific to the current node (Fig.\ref{fig:canvas-UI}.\textcircled{\raisebox{-0.5pt}{5}}, Fig.\ref{fig:node-UI-menu}.\textcircled{\raisebox{-0.5pt}{6}}). Similar to the literature analysis on the side panel, this analysis extracts relevant paper sections that connect to the existing idea facet and proposes next step actions. This node-level analysis provides more granular feedback on each idea facet, helping the user iterate on the idea and receive inspirations for expanding their ideas based on prior works.
The researcher finds the suggestion to evaluate AI character creation across different writing styles compelling. She drags this suggestion to create a chain of ``Evaluation Method'' and ``Contribution and Impact'' nodes, and links them to existing idea nodes. (Fig.\ref{fig:canvas-UI}.\textcircled{\raisebox{-0.5pt}{5}}).

At this point, she wonders how previous works conducted their human evaluations, so she types a targeted literature query in the \texttt{``Q\&A with AI''} interface and receives a response (Fig.\ref{fig:lit-review-panel}.\textcircled{\raisebox{-0.5pt}{3}}). \projname{} synthesizes the response based on the evaluation methods it summarized from each paper. The response lists out the human evaluation method from the collected papers: one paper included an interview with 20 writers on their preferences for AI versus human support, while another paper developed a web application and collected feedback from writers through a ``finish each other's story'' approach, among others.
The researcher notes down these literature-driven insights as she progresses in her project ideation.

\subsubsection{Generating research briefs by exploring different node compositions.}

Using \projname{}'s tools, the researcher refines node content with AI suggestions, deepens her literature understanding through summary and Q\&A capabilities, and broadens her research ideas by exploring alternative approaches via distinct idea facets, ultimately resulting in a more well-defined project. After decomposing her initial idea into different facet nodes and generating variations for each, she feels she has sufficiently explored the problem space. She then selects \textit{a path of idea facet nodes} that encapsulates the research project to generate a \textit{research brief }on the right-side panel. The brief compiles these facets into a cohesive document, with the corresponding nodes highlighted (Fig.\ref{fig:research-brief-panel}).
The researcher shares the research brief with her collaborators to further discuss this project idea. She can also return to the canvas to develop multiple clusters of ideas, creating multiple research briefs to continue ideation. The canvas space, then, functions as a mind map and preserves the ideation context within this broad topic.

\subsection{Node-based Idea Canvas}
\projname{} is built around a node-based canvas, where each node represents a distinct idea facet. 
We specifically define four types of idea facets: \textit{Problem Description and Research Question}, \textit{Proposed Design and Solution}, \textit{Evaluation Methods}, and \textit{Contribution and Impact}.
These idea facets are chosen as prior works have commonly identified them as critical components that represent the research idea in various formats, such as research paper \cite{Mack2012HowTW, Busse2017HowTW, Kallet2004HowTW}, research proposal \cite{Sudheesh2016HowTW}, or abstract \cite{Andrade2011HowTW}, across different scientific domains.
We omitted other potential facets such as evaluation results, or limitations as they cannot be concretely described in early research ideas without execution.
Users can select the different idea facet types via a drop-down-menu (Fig.\ref{fig:node-UI-menu}.\textcircled{\raisebox{-0.5pt}{1}}) as they see fit. 

The node representation decomposes the research idea into a digestible format for iteration and evaluation.
Each node on the canvas offers specific functionalities designed to facilitate interaction. Users can freely edit the node title and content (Fig.\ref{fig:node-UI-menu}.\textcircled{\raisebox{-0.5pt}{2}}).
They can also link nodes together to form logical connections between them.
Each edge is color-coded to represent the strength of the connection (color gradient: red=weak, green=strong) and can be expanded to display text suggestions to enhance the coherence between the connected nodes (\S\ref{prompt:edge-generation}).
\revise{Users can manually create idea facets by right-clicking on the canvas.
Or they can use \projname{} to generate new idea facets by adopting node suggestions (Fig.\ref{fig:node-UI-menu}.\textcircled{\raisebox{-0.5pt}{4}}\textcircled{\raisebox{-0.5pt}{6}}) or manually selecting a facet type (or the same type via ``Generate Alternatives'') at the node menu (Fig.\ref{fig:node-UI-menu}.\textcircled{\raisebox{-0.5pt}{5}}).
A new node is generated based on the context of the current node and any directly linked nodes. The user can use an additional prompt to steer the generation.
Informed by our formative study finding, \projname{} allows researchers to start their ideation with any node type, giving the user control to flexibly navigate the ideation process to diverge, converge, and iterate on any idea facets based on their needs.}
\projname{} also features semantic zooming to hide node details when users zoom out.
The tree-like vertical node structure 
reflects the progression of a research concept from initial stages to a more refined, comprehensive research brief.

\begin{figure*}[h]
    \centering
    \includegraphics[width=\linewidth]{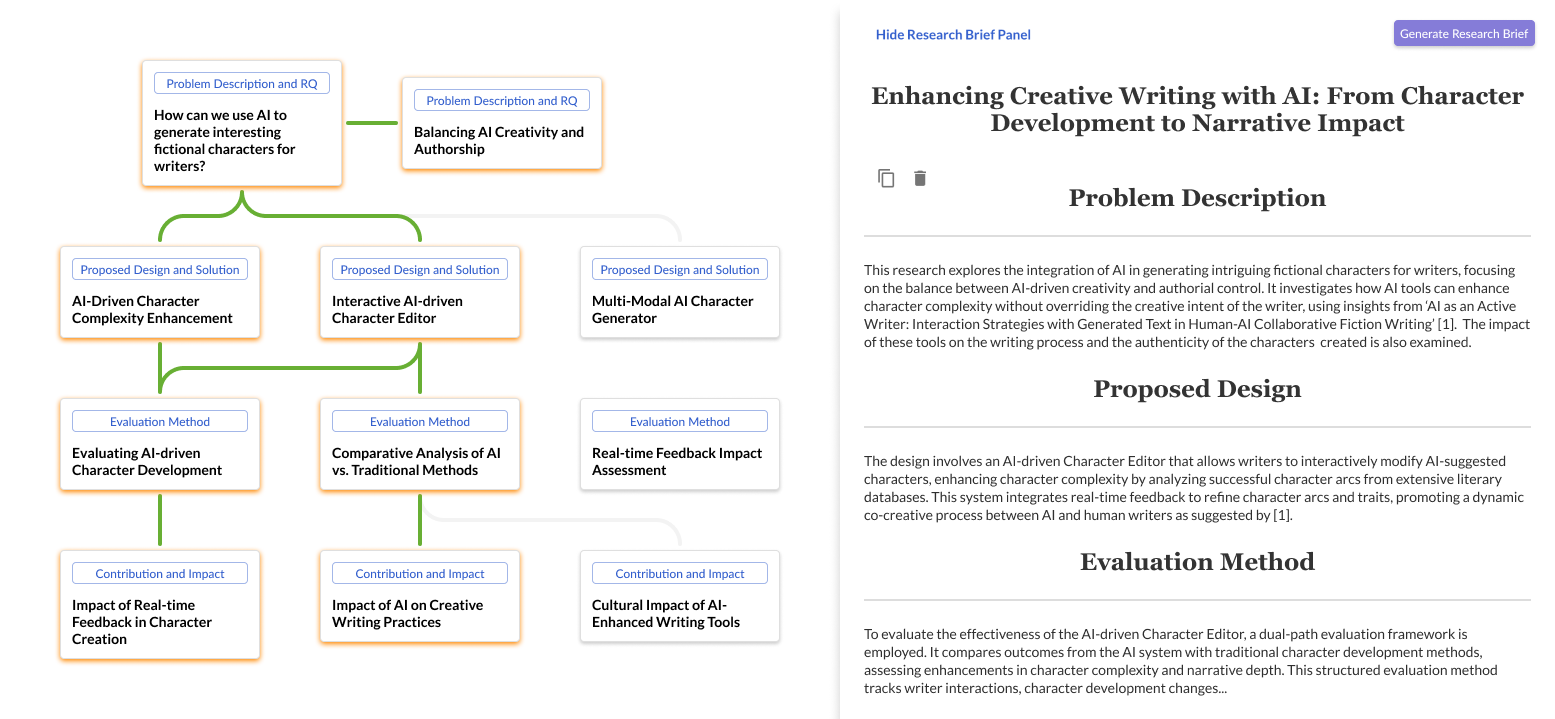}
    \caption{
        \textbf{Research Brief Panel.}
        \textnormal{The user can select multiple nodes on the canvas and generate a research brief on the right panel. The research brief takes content from idea facets and connections and transform them into coherent writing. When focused on the brief, the included nodes and edges are highlighted on the canvas. The user can create multiple research briefs in separate tabs.}
    }
    \label{fig:research-brief-panel}
\end{figure*}

After sufficient idea expansion and exploration, users can select multiple nodes and generate a research brief 
(Fig.\ref{fig:research-brief-panel}, \S\ref{prompt:brief-generation}). This brief integrates the content of each selected node while maintaining the hierarchical connections between idea facets, ensuring that the decomposed ideas and their relationships are preserved. By doing so, the system helps users translate structured representations of ideas into a text-based one for further development or presentation.

\subsection{Methods for Generating Literature-Grounded Feedback}
To provide contextualized feedback for the user's ideas like shown in the scenario, \projname{} implements an LLM-powered system that grounds its generation based on the collected papers (DG3).
The system allows users to construct a paper collection through a search bar (Fig.\ref{fig:lit-review-panel}.\textcircled{\raisebox{-0.5pt}{1}}) connected to the Semantic Scholar API\footnote{\url{https://api.semanticscholar.org/api-docs/}}, which retrieves the details and metadata (title, authors, TL;DR, abstract, etc.) of the paper results based on a search query.
The search result contains real papers from a open scientific literature dataset.
Once the user collects an initial set of papers, they can use the ``Recommend Paper'' function (Fig. \ref{fig:lit-review-panel}.\textcircled{\raisebox{-0.5pt}{2}}), which uses the Paper Recommendations API\footnote{\url{https://api.semanticscholar.org/api-docs/recommendations\#tag/Paper-Recommendations/operation/post_papers/}} to suggest more relevant papers based on a literature knowledge graph using the existing collection as positive examples, as per \cite{Cohan2020SPECTERDR}.

To ensure that \projname{} generates useful suggestions grounded by the collected papers, each paper added by the user goes through additional processing. The system first scrapes the paper PDF file (if open-access) and extracts the full text of the paper using GROBID \cite{GROBID}, an open source library specialized to extract information from technical and scientific publications.
We then provide the full text of the paper and perform retrieval-augmented generation (RAG) using \texttt{gpt-3.5-turbo} to query and summarize specific facets of the paper, including the research question, the proposed design and solution, the evaluation method, the contribution and impact, and the limitations and future works (\S\ref{prompt:paper-processing}).
If the paper full text is not available, the system falls back to the abstract and TL;DR summary to use as context.
We use a lower-tier LLM for RAG with paper-length text to reduce cost and response delay.
\revise{To mitigate LLM hallucination, we delegate the task of finding related paper to their ideation to the researcher, with the support of algorithm-based search engine (i.e. Semantic Scholars) that yields real paper information. 
The user is able to collect any interesting papers relevant to their ideation topic, including works from their own discipline or other domains.
\projname{} only processes and summarizes paper information from the list of papers added by the user using \projname{} and never refer to papers beyond this collection to avoid false source of information.}

Based on the paper collection enhanced with summaries of specific facets, \projname{} dynamically incorporates relevant papers or facets to guide generations for idea refinement (DG1) and explore the information space (DG2).
\projname{} generates LLM suggestions to scaffold the research ideation process at two abstraction levels.
Overall, at the node canvas level, \projname{} allows the user to generate a one-paragraph literature review summary based on all the collected papers (Fig.\ref{fig:lit-review-panel}.\textcircled{\raisebox{-0.5pt}{4}}, \S\ref{prompt:lit-review-summary}).
The system additionally analyzes the literature using the current ideation progress, including all the node and edge contents on the canvas as context (\S\ref{prompt:lit-review-anlaysis}). The literature analysis generates several comparisons between collected papers and idea facets, each identifying the key section (i.e. facet) in a particular paper and highlighting the connection to the idea facets (Fig.\ref{fig:lit-review-panel}.\textcircled{\raisebox{-0.5pt}{5}}). 
\projname{} also provides actionable suggestions for the user to iterate on existing idea facets or generate new ones, such as dragging the suggestion element onto the canvas to automatically generate a chain of nodes that executes on the suggestion (\S\ref{prompt:node-generation}). Additionally, the Q\&A interface allows users to ask targeted questions about specific details in the collected papers or inquire about ideas represented on the canvas (Fig.\ref{fig:lit-review-panel}.\textcircled{\raisebox{-0.5pt}{3}}, \S\ref{prompt:ai-chat-qa-response}). 

As for the idea nodes, \projname{} offers three types of node suggestions: idea iteration, idea alternatives, and idea expansions (Fig.\ref{fig:node-UI-menu}.\textcircled{\raisebox{-0.5pt}{3}}, \S\ref{prompt:node-suggestion}).
To help refine and iterate on the user's idea (DG1), the system asks clarifying questions to prompt deeper reflections and more detailed descriptions, encouraging users to address potential challenges based on existing works.
To facilitate idea exploration (DG2), \projname{} suggests alternative variations for the current idea facet to broaden the user's exposure to research questions, designs, evaluation methods, and more. It also recommends ways to expand \textit{between} idea facets, such as proposing solution designs for a given research question or deriving research questions linked to an evaluation method. Users can manually adopt these suggestions or let \projname{} generate them automatically, with options to customize the prompts. For each node, the system generates a node-specific literature analysis based on its content and type, leveraging relevant collected papers and suggesting actions for further iteration or expansion (Fig.\ref{fig:node-UI-menu}.\textcircled{\raisebox{-0.5pt}{6}}, \S\ref{prompt:node-lit-review}).
Through these methods, \projname{} dynamically grounds LLM generation in the user's collected literature at both the canvas and node levels. This provides users with actionable and contextually relevant suggestions for refining, expanding, and iterating on research ideas.

\subsection{Implementation Details}
\projname{} is implemented as a web application. The backend was implemented in Python using Flask. The frontend used the React framework in Typescript. We used LangGraph \footnote{\url{https://github.com/langchain-ai/langgraph}} to instantiate an LLM instance with memory of past interactions and generations. \revise{We used \texttt{gpt-4-turbo} with default parameter values supported by LangGraph (\texttt{temperature$=0.7$}). However, \projname{} is designed to be model-agnostic and our key contribution is on the human ideation workflow enhanced by incorporating LLM assistance. We adopted common prompt strategies like chain-of-thought and few-shot prompting, which yielded higher quality results from our testing.} All LLM prompts used in the system are included in Appendix \ref{section:llm-prompts}.

\section{Study 1: Comparative Lab Study}
To evaluate the usability and utility of \projname{}, we first conducted a within-subjects laboratory study where participants developed research ideas using \projname{} and a baseline system. We formulated the following research questions:

\begin{itemize}
    \item \textbf{RQ1:} How does the node-based canvas interface allow users to explore and connect different idea facets? 
    \item \textbf{RQ2:} How does the externalization of idea facets help the expansion and refinement of initial research ideas?
    \item \textbf{RQ3:} How do LLM-generated suggestions grounded on collected papers help the user understand the literature space and evaluate their research ideas?
    \item \textbf{RQ4:} How does the user experience of research ideation using \projname{} compare to using a standard LLM-based tool?
\end{itemize}

\subsection{Study Design}

\subsubsection{Participants.}
We recruited 20 researchers (11M, 8F, 1 non-binary) who are pursuing or have completed their Ph.D. degrees (i.e., graduate students and postdocs) in Computer Science through posting recruiting messages in Slack channels of several research institutes. 
We targeted scholars who were actively engaging in research ideation and project development.
In total, twenty participants were recruited, of which 12 were in the field of HCI and 8 in the field of NLP. Thirteen participants have 2--5 years of research experience, six have more than 5 years of research experience, and one has less than 1 year of experience. 
\revise{From the entry survey, all participants described literature review and discussion with peers/advisors as a part of their research ideation process.
Twelve participants mentioned writing a research plan or proposal as a part of their ideation, utilizing different tools for documenting ideas and related work notes (e.g. Google Doc, Notion, Obsidian, etc.)}
Participant demographics can be found in Appendix \ref{a:lab-participants}.

\subsubsection{Conditions.}
\label{s:evaluation-condition}
The study adopts a within-subjects design with two conditions: \projname{} and a strong baseline. 
Simulating a common practice reported by participants in the formative study, the baseline system implements a word editor\footnote{\url{https://quilljs.com/}} where participants can develop their research ideas. 
To further control the task goal, the baseline editor provided an initial writing template containing section headings corresponding to the idea facet categories in \projname{}.
The baseline condition also contained a subset of features in \projname{}. Specifically, a literature review panel (Fig.\ref{fig:lit-review-panel}.left column) with identical paper search and recommendation features to support literature understanding.
The baseline system also provided LLM assistance through the same Q\&A interface. 
The user can ask targeted inquiries regarding the collected papers or free-form questions about their research ideas similar to using tools like ChatGPT. 
The two systems share identical underlying LLM setup and system prompt. The LLM in the baseline is supplemented with the same paper metadata and the writing editor content as context. 
Additionally, the baseline contains an \texttt{``AI Assist Writing''} button in the editor, which will generate a revised research brief based on the user's writing and attached it at the end of the editor.
This function uses the same prompt (\S\ref{prompt:brief-generation}, but with editor content as context) as the \texttt{``Generate Research Brief''} function in \projname{} that transforms idea facet nodes into a coherent writing. 

\subsubsection{Tasks.}
Each participant was asked to develop research ideas based on two provided broad topics. For HCI researchers, we identified a main theme of human-AI interaction and two task topics: \textit{``AI-augmented Tool for Creativity''} and \textit{``Human-AI Collaboration in Data Anlaysis''}. For NLP researchers, we chose the theme of LLM evaluation, with the two task topics being \textit{``LLM-as-a-Judge''} and \textit{``Biases in LLMs''}.
The task themes cover a wide range of concepts and specific domains to provide sufficient freedom for researchers in different areas to develop research ideas. 
We did not let participants ideate on their existing research projects as there might be variance in the stage of ideation between projects and participants.

\subsubsection{Procedure.}
We control for the learning and ordering effects by counterbalancing the system condition and task topic order ($2\times2$) independently for the HCI participant group ($n=12$) and the NLP participant group ($n=8$).
The participants were first given a brief introduction on the goal of the study and a task instruction sheet (Appendix \ref{a:task-instruction}).
For each task, they first watched a short tutorial of the assigned system (\projname{} or baseline), then spent 30 minutes to develop their research idea on the assigned task topic. When there were 5 minutes left, participants were reminded to make final changes to their idea and converge on a research brief as the task deliverable.
After each task, participant filled out a survey to evaluate the quality of their research idea and record their perception about the use experience and the task workload.
At the end of both tasks, we conducted a semi-structured interview where participants reflected on their overall experience and compared the ideation process using both prototypes.
The study was conducted remotely using video conferencing software. Each session lasted around 90 minutes, and participants were compensated for \$75 USD. The study was approved by our internal review board.



\subsection{Measures and Analysis Procedure}
We collected both users' perception and behavioral data to conduct a comprehensive analysis on the usage of \projname{}. Quantitatively, we analyzed the post-task survey responses. The survey consists of 7-point Likert scale questions on the user's perception towards their level of literature understanding, idea refinement, and idea expansion they were able to achieve during the task. Additionally, the survey includes questions from NASA-TLX \cite{hart2006nasa} on perceived workload, trust, and five questions asking the participants self-assess the quality of their final research idea in five dimensions: novelty, impact, specificity, feasibility, and relevance, defined by prior work \cite{douglas2006identifying}. 
We pre-selected four key metrics from the survey for statistical testing using a MANOVA test to account for multiple comparisons. The four key metrics were selected because they were either common and important aspects when evaluating research proposals (i.e., \emph{impact} and \emph{feasibility}) or closely related to our design goals based on our formative study (i.e., \emph{confidence in literature understanding} and \emph{sufficiently explored and evaluated alternatives}).
The full list of survey questions and results are presented in Appendix \ref{a:survey-response}.
In addition, we analyzed the interaction logs from the sessions to measure the quantity and types of actions taken (e.g. add a paper to collection, get AI suggestion, user typing, etc.) and capture behavioral differences.
We used Wilcoxon signed-rank tests for pairwise two-condition comparisons for independent Likert scale survey questions (Fig.~\ref{fig:survey-res}) and T-test for temporal data.
For the qualitative data, the first author performed thematic analysis on the interview transcripts and iterated on the themes with the other authors through discussion. We report the synthesis of our themes, denoting HCI participants from H1--H12 and NLP participants from N1--N8; our full themes can be found in Appendix~\ref{a:codebook-lab}.

\subsection{Results}

\subsubsection{\textbf{\projname{} helped users better explore alternative ideas and navigate the idea space (RQ1).}}
\label{s:findings-more-exploration}
\begin{figure*}[t]
    \centering
    \includegraphics[width=\linewidth]{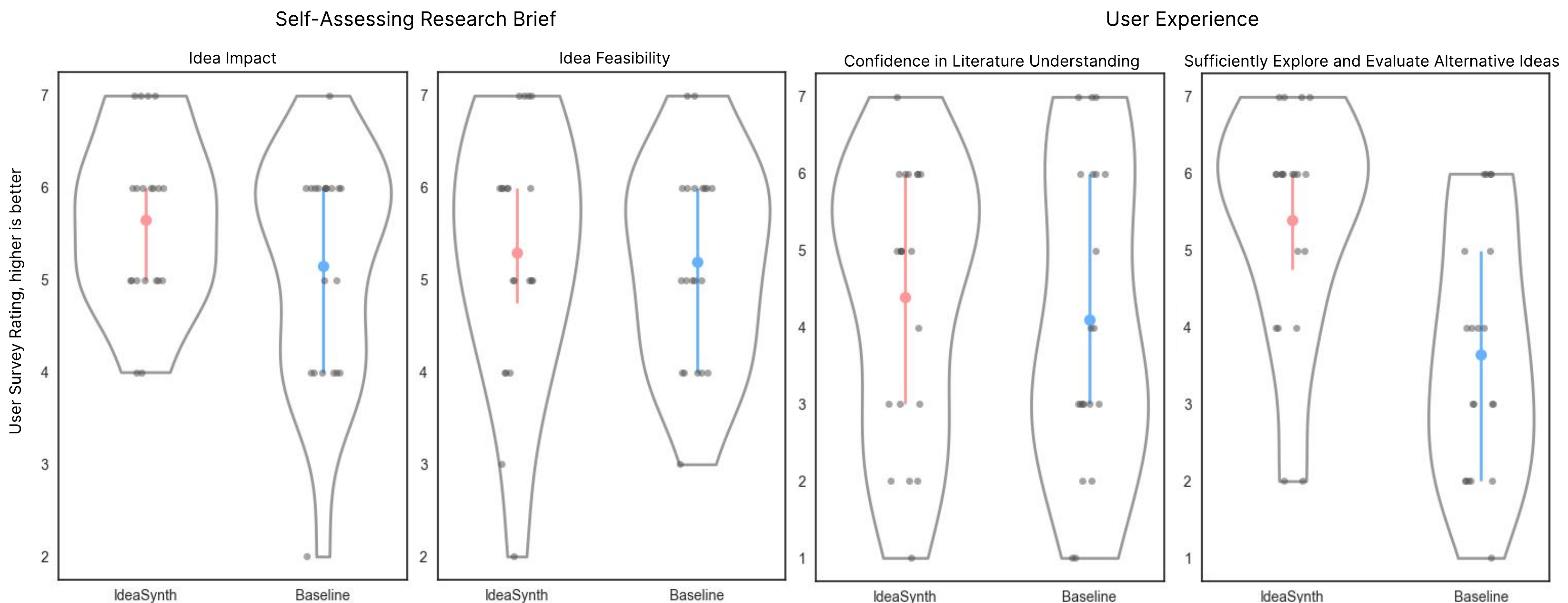}
    \caption{
        \textbf{Key Survey Responses}
        \textnormal{Participants generated research briefs after 30 minutes of ideation with each of the two systems. Participants self-assessed their research brief and rated their user experience and perceived workload after each task. We pre-selected four key measures and conducted a MANOVA test to correct for multiple comparisons, identifying a statistically significant difference ($manovaF=3.30, p<0.05^{*}$) between \projname{} and the baseline on the combined dependent variables of Impact, Feasibility, Literature Grounding, and Sufficient Exploration. We separately report full survey questions and results in Appendix \ref{a:survey-response}.}
    }
    \label{fig:survey-MANOVA}
\end{figure*}

Based on the survey response, participants felt they more sufficiently explored and evaluated alternative versions of their research ideas in \projname{} ($mean=5.40$, $SD=1.50$) than the baseline ($mean=3.65$, $SD=1.60$, $p<0.01$), as shown in Fig~\ref{fig:survey-MANOVA}. H3, H10, H12, and N3 stated that the baseline system, due to its editor-like interface, did not encourage exploring alternative ideas but pushed them to develop their initial idea in depth. However, the lack of exploration could lead to idea fixation (H3).
In contrast, fourteen participants mentioned that the support for suggesting and generating alternative nodes in \projname{} led them to explore more \revise{adjacent and complementary} idea facets. H1, H2, H6, and H11 attributed the enhanced process of exploration to the structured layout with horizontal levels of nodes for each idea facet, visualizing the connections between ideas.
N6 thought \projname{} help them \textit{``find the connection between two nodes and this feature can help you discover something you didn't recognize before.''}
\revise{H7 demonstrated the process of expanding and developing an abstract idea into a concrete research brief. They initially created a ``Contribution and Impact'' node with a general idea to help people understand their health data. Utilizing node-level literature suggestions, H7 generated multiple separate paths of nodes with other complementing facets. Then, H7 connected idea facets that were interesting to them to synthesize their proposed solution in a research brief, including designing an AI health assistant across multiple sensing devices and integrating staged-based model to structure interactions across data collection, integration, reflection, and action.}

The ability to evaluate multiple generated idea facets based on collected papers broadened participants' perspectives (H1, H12). H1 claimed that \textit{``since [\projname{}] actually generated new ideas and suggested some papers based on some of my prompts, it actually broadened my understanding on the area.''}
Sometimes, the generation of alternative ideas inspired users \revise{to extend their ideas with details} they haven't considered. 
H8 reflected that through using \projname{}, their research idea \textit{``grew into providing more details or sometimes, an angle that I hadn't thought about.''} 
Participants also reported going through a converging stage during idea exploration. N3 felt like the alternatives generated by \projname{} were \textit{``more out of my scope and out of my expectations. So that's good. Not necessarily means that those ideas are good but still it elicits some new thoughts in me. That's also indirectly helping me to find some good directions.''} H2 expressed that they were more confident in their own idea after generating and evaluating alternatives, pruning irrelevant nodes, and converging on the scope of the idea. 
For participants who ideated on unfamiliar topics (H9, H4), the AI suggestions helped them explore and better understand the information space. 
\revise{For example, H9 began an ideation session on using AI for painting, an unfamiliar topic for them. By searching related works and reviewing literature summaries, they gained insights into how AI-driven approaches influence traditional artistic processes. H9 organized their ideas into two branches, exploring both the benefits and risks of AI-assisted painting. Using insights from the collected papers, they generated ``Problem Description and RQ'' nodes, evaluated artists' perceptions of AI, and iteratively refined their exploration. This process culminated in a tree-structured canvas, from which H9 selected key facets to craft the final research brief.}

Despite the benefits in promoting exploration and ideation, \projname{} has some limitations. 
Four participants (H1, H12, N5, N6) noted that the system's node-based layout displayed dense information that increased their mental workload.
For example, H12 expressed that navigating through numerous suggestions required extra effort to sift through and prioritize relevant ideas and occasionally felt \textit{``a bit overwhelming''}. 
On the other hand, this sentiment was not reflected in the survey results ($\projname{}: mean=4.35, SD=1.73, Baseline: mean=4.55, SD=1.64$, Fig.\ref{fig:survey-res}),
suggesting that balancing the costs and benefits of the node-based layout and combined with other features of \projname{}, the overall mental workloads were comparable. 
Finally, one of the participants (N5, an algorithm researcher), preferred the traditional writing-centric approach, such as using physical pen and paper, which might be specific to research areas like algorithm design where the tactile process of sketching out ideas and deriving mathematical equations felt more intuitive. 
These observations suggest that, while \projname{} supports ideation effectively, its design could be optimized to balance idea generation with cognitive load, as well as consider researchers' needs in different domains. We further discussed this idea in the Discussion and Future Work section.

\subsubsection{\textbf{\projname{} helped users expand and iterate on ideas (RQ2).}} 
\label{s:findings-expand-iterate}
Almost all participants ($n=17$) reported that they were able to more effectively expand their initial research idea with additional details with \projname{}. This was supported by the survey responses ($\projname{}: mean=6.05$, $SD=0.89$, $baseline: mean=4.45$, $SD=2.04$, Fig.\ref{fig:survey-res}).
Many participants ($n=11$) attributed this to being able to break down their ideas into multiple facets using the node representation. 
Upon connecting facets to compose them, H7 felt more motivated to expand their idea: \textit{``once you have that initial idea to start with, that's when I can see potential for a new node, 
... I can imagine each node being a research question and other nodes which are connected to it through edges,...[it's] motivating [as] that's when I would actually pay attention to the connections.''}
\revise{One instance that demonstrates \projname{}'s assistance in expanding and iterating on ideas is H8's exploration of using AI to generate fictional characters' key story moments based on a general description. By evaluating \projname{}'s node suggestions, H8 iteratively expanded the scope of their research idea to focus on elevating narrative flexibility in creative writing, envisioning an interactive character editor powered by LLM-driven complexity enhancement. Additionally, alternative nodes generated by \projname{} helped H8 identify and integrate the balance between AI creativity and human authorship as a critical ``Problem Description and RQ'' node—an aspect of the research they had not initially considered but later incorporated into their research brief.}

\revise{Similarly, N6 leveraged \projname{} to refine their initial idea of reducing cultural bias in LLMs. Building on \projname{}'s suggestions, they connected two ``Proposed Design and Solution'' facets: one focusing on semantic data augmentation and the other on a cultural bias feedback loop. Through \projname{}, N6 identified the opportunity to quantify the effectiveness of data augmentation by integrating an iterative feedback loop that combined both facets. Inspired by this connection, N6 iterated on the node contents to detail a human evaluation system for dynamically tuning data parameters based on feedback, adding depth to their research direction.}
Comparing \projname{} to the baseline system, H3 said that \projname{} \textit{``was more targeted in a sense that they [ideas] were separated into the specific things that I had to include in the research brief so I can iterate on one component much easier,''} whereas the baseline \textit{``just provided general summary that was very general.''}

The majority of participants ($n=13$) also found that the AI suggestions for each idea facet in \projname{} were helpful for iteratively refining their ideas.
N2 felt that the suggestions \textit{``motivated [them] to think of more details''} \revise{from their original idea description} via clarifying questions.
H5 experienced \textit{``more control on iteration... after I learn more about this topic, I can add more details... it's definitely easier to focus on a specific aspect, easier to iterate on that specific node.''} 
H9 also reflected that the node connections were helpful for iterating on an idea as the suggestions are \textit{``constantly poking holes and connections, and finding ways to make more connections...while also strengthening my proposal.''}
In contrast, participants felt that AI suggestions in the baseline system were more like summarization rather than ideation (H8, N4, N8), similar to interacting with tools like ChatGPT (H5, H8, N4, N6), despite using the same underlying LLM and system prompt in both conditions.

\subsubsection{\textbf{Similar literature search behaviors but \projname{}'s facet-specific summaries and suggestions can surface unexpected connections to prior work (RQ3).}}
\label{s:findings-literature}
In general, participants ($n=12$) found the literature review panel in both conditions helpful for exploring related works and gaining an overview of the papers to provide a foundation for research ideation. 
We did not find a significant difference in users' confidence in understanding the literature in the post-task survey across two conditions.
The behavioral logs also showed that participants spent similar percentages of their task duration searching and reading paper information ($\projname{}=9.75\%$, $Baseline=11.7\%$), and we did not identify a significant difference in the number of papers collected ($\projname{}=5.60$, $Baseline=6.05$, $p=0.63$).
This is unsurprising as both systems included the same literature search features.

However, some participants ($n=7$) found \projname{}, with the addition of literature summary and analysis generation at both the canvas-level and node-level, more helpful compared to the baseline.
\projname{} summarized the collected papers based on the defined idea facets.
H7 found it helpful to seek literature support for ideating specific facets in their ideas: \textit{``I think that's what I need when I'm brainstorming or ideating. I am confident enough in my methods to not need help...but I think even then this is summarizing motivation...that was pretty helpful.''}
N4 commented that \textit{``the option to generate a summary of various paper and...being able to see some connection between the two papers is very interesting and not something that I saw before.''}
Similarly, N7 felt that \textit{``[\projname{}] was much better at not just giving you a summary of papers
... it's... connecting the dots for you.''}

\projname{} additionally presented relevant paper information for individual nodes. 
H4 commented that \textit{``just being able to explore different types of suggestions was super useful...be able to toggle through and see how it was bringing up some of the literature that I had collected.''}
H4 felt that the literature-grounded AI suggestions helped them maintain context of relevant works and prevented idea fixation: \textit{``It's mostly to offer suggestions and things that wasn't immediately thinking about. It just helps to prevent myself from pigeonholing because it will point out different papers that I couldn't just think of referencing immediately that strengthen the research idea...Or talking about related work that I could build upon instead of starting from scratch.''}

Yet, some participants ($n=6$) shared that the features in \projname{} and the baseline do not replace the literature review process of actually reading the papers.
H4 commented that \textit{``I trust the [literature] summaries, but I don't trust that it's done an exhaustive search of papers yet.''}
Participants found the LLM-generated suggestions based on collected papers can sometimes still be too broad (H6, H10), not entirely trustworthy (N7, H11), and over-reliant on the scope of the collected papers, placing the effort on the users to find a representative initial set of papers (N3, N7, H10).
We note that the main focus of \projname{} is not on literature discovery or understanding.
We incorporated basic paper search and summary functionalities as literature review plays an important role in research ideation. Future work can explore how to better integrate literature review with research ideation.

\subsubsection{\textbf{Participants gained increased benefits from AI support in \projname{} and spent less time writing from prior knowledge (RQ4).}}
To identify the differences in user behavior between the two conditions, we analyzed the interaction log from the user study sessions, where we recorded timestamped user or system actions.
We recorded overt actions such as clicking on buttons to trigger system features (e.g. ``Get AI Suggestions'', Q\&A, search paper), users' typing traces (e.g. writing research ideas in node content, or in the editor document).
We further grouped actions as user-driven actions, such as writing and reviewing literature, or AI-assisted actions, such as generating a response to the user's prompt, which is present for both conditions, or generating new nodes in \projname{}.
We find that the participants utilized the AI assistance more frequently in \projname{} than in baseline.
On average, they spent 36.8\% of their time during each task on AI-assisted actions in the treatment condition and 12.9\% in the baseline condition. This is expected as \projname{} integrates more LLM-based features to provide suggestions and scaffold idea expansion and exploration throughout the ideation process.

In contrast, participants spent a significantly larger portion of time writing in the baseline without utilizing features from the system.
On average, they spent 19.2\% of the time for each task on writing actions in \projname{} (writing node title, node content, editing generated research brief) and 64.8\% of time writing (writing in the document interface, editing AI-revised research brief generated using \texttt{``AI Assist Writing''}) in the baseline condition.
Participants' uninterrupted writing sessions were also longer in the baseline. Analyzing at the action level, an uninterrupted writing action (i.e., no switching to other actions) in baseline lasted on average 124.3 seconds, compared to 51.5 seconds in \projname{}. However, participants engaged in similar total numbers of discrete writing actions ($\projname{}=121$, $Baseline=125$), making smaller writing edits for each decomposed idea facet in \projname{}, but wrote more extensive passages in the baseline.
While spending more time on writing is not necessarily a hindrance to ideation, the document interface in the baseline did not provide users with adequate support for simultaneous divergent exploration of multiple ideas, according to some participants ($n=9$). 
H10 remarked that they \textit{``mostly just used the [baseline] to basically prosify the idea that [they] already had, or maybe flesh it out in a little bit more detail, but not necessarily as much to consider alternatives.''}
Similarly, H3 recalled that while using the baseline system, they were \textit{``kind of fixated on a design and it's too much work [to enter] in the whole thing into the assistant [Q\&A interface] and ask it for different alternatives. So I just click on the `AI assist writing' to make it to polish the research brief instead of trying to do more ideation.''} 


\subsubsection{\textbf{Participants felt \projname{} improved task success, but further study is required to evaluate research idea quality.}}
Participants rated their perceived level of success in accomplishing the task slightly more highly in \projname{} ($mean=5.15$, $SD=0.81$) than the baseline ($mean=4.55$, $SD=1.23$, $p<0.05$ using independent Wilcoxon signed-rank test, Fig.\ref{fig:survey-res}).
In the combined dependent variable, which included Impact and Feasibility as key measures for research brief quality, we identified a significant difference (Fig.\ref{fig:survey-MANOVA}) for \projname{} over baseline. 
However, measuring participants' ratings on the five dimensions---Novelty, Impact, Specificity, Feasibility, and Relevance (full definitions in Appendix \ref{tab:survey_labels})---independently,
we did not identify a significant difference between conditions for any metrics.
This might be due to the nature of uncertainty in the research ideation. The comparison of ideas might also be affected by the anchoring effect of the initial insight participants chose to develop.
N5 expressed difficulty in evaluating their ideas in both conditions: \textit{``I think the task [of rating idea quality] is honestly a little hard...the questions I answered about feasibility or specificity, it wasn't about the system that I used, it's just that particular idea that I came up with.''}

We wanted to capture participants' initial perceptions of the research ideas they developed in a controlled lab setting.
However, evaluating research ideas and predicting research outcomes are difficult \cite{Si2024CanLG}, even when reviewing complete paper submissions \cite{Shah2022AnOO, Arous2021PeerGT}.
Qualified experts are difficult to recruit and can still face difficulties in judging the quality of a research idea as evaluation criteria can be perceived highly subjectively \cite{Si2024CanLG, Beygelzimer2023HasTM, Simsek2024DoGP}.
To further investigate the naturalistic usage and ecological validity of \projname{}, we conducted a deployment study described in the following section. 
We further discuss the need for an end-to-end study to predict the ideation outcome in the Discussion.


\section{Study 2: Field Deployment Study}
To further investigate how scholars would use \projname{} to develop real-world research ideas organically, we deployed \projname{} and invited participants from the lab study to continue using the tool on their own computers. Participants were asked to use the tool for at least three days, for at least one hour in total usage, and return for a 15-30 minutes follow-up interview. All seven participants (D1-D7; 3 in the field of HCI and 4 in NLP) completed the deployment study and were compensated \$25 USD for their time.
The first author conducted thematic analysis on the interview transcripts\citep{Boyatzis1998TransformingQI,Connelly2013GroundedT} (Appendix \ref{a:deployment-codebook}). Below we report our findings.

\subsection{Deployment Study Results}
\subsubsection{\revise{\textbf{Participants used \projname{} for research ideas in different phases of ideation.}}}
While we focused the design of \projname{} on supporting idea expansion and refinement,
in a field setting, participants tested the system across a wider range of the ideation phase.
Predominantly, participants used \projname{} for two primary tasks: research ideation and literature understanding.
For research ideation, participants were able to apply \projname{} to their research projects in various stages of development.
D1, D3, and D7 explored early insights in \projname{} and expanded them into more concrete structured research ideas.
D3 described their process as \textit{``[like] I was having a discussion with my supervisor, I think we have a rough idea right now... I was trying to do some research ideation because I don't really know anything about [topic]...it [\projname{}] gives me a general thinking flow, there's problem, research question components all the other stuff. So it gives me a complete research ideation process and the proposal.''}
Similarly, D7 found the tool \textit{``convenient because I'm sort of in the midst of a brainstorming phase for a different project. So mostly I just used it to sort of help me come up with ideas to think about different ways of framing, research questions, different methods to use.''}
Some participants ($n=3$) externalized their existing ideas using the node-based canvas and examining the connections between idea facets to reflect on their existing ideas more clearly.
D4 applied \projname{} on an idea where they already drafted the paper on and \textit{``already have a structure of how the research should go. And these texts [on the idea facet nodes] are filled by myself. And this time I focused on using the relationship [node edges] and I feel pretty surprised. I think this one really works pretty good. It gives me some suggestions...and it tells me the strengths of the connection.''} 
Despite further developing a more mature research idea, D4 found benefits using \projname{}: \textit{``my idea wasn't this complex at the very beginning. It really helped me to make my ideas completed with many branches and I was able to derive something pretty useful.''}
D2 also utilized the tool for an ongoing project: \textit{``I was trying to use the tool to help me with related work and how I might design evaluations.''}
On the other hand, most participants ($n=5$) found utility in contextualizing their research ideas using \projname{}'s literature search and summary functionalities.
For instance, D6 commented \textit{``because I have a paper deadline, I was trying to use the tool to more strengthen, like contextualize my paper in the HCI literature
.''}

\subsubsection{\revise{\textbf{\projname{} fits into researchers' ideation workflow and can be used long-term.}}}
We also collected participants' response on how \projname{} can fit in their research workflow and what some envisioned use cases are.
Most participants ($n=5$) expressed interests to continue using \projname{} for research ideation. D4 felt that the workflow of the tool was \textit{``fun to use''} and that \textit{``one of the biggest contributions of this tool is it brought the barrier in my mind to start a research or start thinking or initiating this process way easier...and this fun or kind of excitement of using this tool to exploring project ideas makes me more like to do it.''}
D3 and D7 commented that even without the AI features, they would consider using \projname{}'s representation of idea facets to organize and structure their ideas to stimulate thinking.
Over the course of the entire research process, D5 suggested they could use the tool to perform \textit{``periodic checkpoints in the projects' lifespan where you force yourself to think at a more high level... and how this fits into the larger narrative.''}
D1, D2, and D6 also see potential in using \projname{} as a form of collective ideation tool, where researchers can share and iterate on the ideation canvas with collaborators.

\subsubsection{\revise{\textbf{Participants viewed \projname{} as an assistant to support and advocated for human involvement during ideation.}}}
Regarding the role of LLMs in research ideation, most participants ($n=4$) expressed some level of concerns on the trustworthiness of LLM-generated literature summaries and research ideas.
D1 experienced a conflict for adopting the literature findings presented by \projname{} directly and continue with ideation or verifying them by reading the paper, sharing concerns on potential \textit{``misinterpretation''} of collected papers.
D4 felt that in general, \textit{``we cannot 100\% believe that something that has been generated and click next''}, citing worries about hallucination.
D2 echoed the concern on generated ideas. Despite using ChatGPT in their research, they \textit{``don't ask LLM to generate ideas for me...I do kind of worry about the novelty variation from the LLM. I didn't trust that it can actually evaluate the novelty.''} 

For all participants ($n=7$), an ideal workflow collaborating with \projname{} or AI systems in general consists of using LLM tools to assist and support the researchers in their ideation process.
D6 expressed that while LLM can be useful for exploring ideas, to make decisions and evaluate the ideas, \textit{``human need to be more involved rather than just 99\% using the tool.''}
Likewise, D7 reflected that \textit{``it's [\projname{}] most useful just as a way to sort of stir the mental pot. Even if I don't find myself directly wanting to copy some of the suggestions...I like having that explicit graph structure and sort of explicitly generating alternatives...it sorts of get things bubbling a bit more mentally and forces you to reconsider things.
''}
We further describe the study implications for future human-AI research support tools \revise{and the risks of using LLM for research ideation} in the Discussion.
\section{Discussion and Future Work}

\subsection{Tailoring AI Support to User Needs across Different Stages of Research Ideation}
Our findings highlight that users desire varying levels of AI support depending on the stage of their research process. Specifically, our participants employed \projname{} in different ways during the pro-longed deployment study, emphasizing the need for flexible control, customizability, and adaptability of AI ideation systems.

In the early stages of research ideation, participants appreciated \projname{}'s ability to suggest alternative ideas and connections, which helped broaden their thinking. 
However, the desire for even more control over AI-generated suggestions was frequently mentioned. Users expressed interest in understanding the rationale behind the system's recommendations, particularly when certain papers or ideas were excluded. For instance, H2 questioned why certain relevant papers were not suggested and wanted more insight into the decision-making process behind AI outputs: \textit{``It [system] puts it [generation] through a funnel... but I'd like to potentially just explore all possibilities.''} This feedback suggests a need for greater transparency and explainability in AI-driven suggestions to ensure users feel empowered rather than constrained by the system.

As an idea becomes more developed, some participants switched to a more hands-on strategy, taking over more agency by selectively engaging with \projname{}'s suggestions to take greater control over their canvas workspace. 
However, H5 and H6 also reported that while user-driven interactions afforded greater control, they also require higher mental workload, which may hinder creative flow (H5), raising potential needs for additional passive AI support to streamline ideation (H6). 
For example, H6 remarked, \textit{``I generally want something that passively helps me rather than me having to call into it to help me.''} This highlights the tension between user-initiated and passive AI assistance, where users seek a balance between ease of use and meaningful involvement in the ideation process.

During the deployment study, participants further specialized in using \projname{}'s features based on their workflow needs. Many found it useful both for literature discovery and ideation but often separated these tasks. D7, for example, preferred to use the node-based system for structuring their research while leveraging the literature review features for understanding and differentiating related work. Similarly, D4 and D5 praised \projname{}'s ability to retrieve high-quality papers and quickly summarize them, using it primarily as a tool to strengthen their understanding of the research landscape.
On the ideation side, participants like D1 and D3 used \projname{} to explore and refine early-stage ideas, adding and removing nodes as their mental models evolved. Others, like D4, applied the tool to mature projects, finding it helpful for expanding and clarifying connections between research facets. The flexibility to adapt \projname{} based on research idea maturity underscored the importance of customization, as participants tailored the system to their specific stages and tasks, whether brainstorming ideas or fine-tuning concepts.
Overall, these findings highlight the need for flexible, user-controlled AI support to align with the dynamic and multifaceted nature of the research process.

\subsection{Supporting User-Driven Research Processes Instead of Full Automation}
There has been increasing interest in employing LLMs to fully automate some or all parts of the research process. Si et al. \cite{Si2024CanLG}, for example, found evidence that LLM-generated initial research ideas are rated as more novel than ideas generated by human experts.\footnote{It is worth noting that most expert ideas were broad initial ideas that expert participants came up with on the spot specifically for the study.} Lu et al. \cite{lu2024ai} introduced a system that automates the entire AI research workflow, from ideation, to experiment execution, to paper writing. While these works show that \textit{AI alone} can match or even surpass the research task performance of \textit{humans alone}, they do not reflect the more common real-world scenario where humans and AI tackle research problems together in \textit{human-AI teams}. 

Our work provides an example of how carefully designed user interfaces and interactions for LLM-powered research tools may be key to enabling researchers to collaborate effectively with AI models. For example, by breaking down ideation into idea facets, researchers could have greater control over how and where AI assistance is employed to further align with their goals during ideation. Participants in our study could easily expand and iterate on variations of a specific idea node within the context of a broader topic (Section \ref{s:findings-expand-iterate}). They were able to use the LLM's breadth of knowledge to surface ideas and details that they had not previously considered (Section \ref{s:findings-more-exploration}), as well as uncover new connections between papers (Section \ref{s:findings-literature}). On top of that, our faceted design also allowed participants to better steer the AI when it completed a task in a suboptimal manner (e.g., participants mentioning that LLM outputs were too generic in Section \ref{s:design-goals}), which is much more challenging to do in fully automated AI research pipelines. 

As we observed in Section \ref{s:formative-c1}, researchers may not always want or need AI assistance throughout the entire life cycle of a research task, but may appreciate it at specific moments. We demonstrate by strategically guiding AI to assist with tasks where it excels, we can unlock new potential for researchers and AI to co-create higher-quality research outcomes than either humans or AI alone.

\subsection{\revise{Implications of Using LLMs for Research Ideation}}
\subsubsection{\revise{\textbf{Risks of Using LLMs for Research Ideation}}}
\revise{While utilizing LLMs enables efficient summary and synthesis of information based on the collected papers and user-written idea facets, the model generation could contain inherent biases that negatively impact the outcome of the research ideation.
Studies have shown that LLMs could exhibit social biases (e.g. gender, race, religion) stemmed from its training data \cite{liang2021socialbias, navigli2023bias}.
Specifically, Dai et al. identified bias and unfairness from the recommendation and results in information retrieval systems integrated with LLMs \cite{dai2024IRbias}.
In the scientific research context, this could lead to misrepresentation of certain publications, findings, or theories.
Moreover, LLMs could potentially skew the research direction by presenting dominant or popular ideas and overlooking more novel or less-explored approaches, which could harm researchers' creativity. 
Another potential risk is hallucination, where an LLM generates content that misaligns with established world knowledge or previous interaction context \cite{zhang2023sirenssongaiocean}.
This could lead to false ideas or fabricated sources being used during the ideation process.}

\revise{To mitigate the above risks, \projname{} grounds its generation on the curated paper collection.
The generated idea facets and suggestions are often distilled from the original context of the collected papers in an effort to prevent hallucination.
In an attempt to prevent misrepresentation of ideas, \projname{} puts the user at the steering wheel when exploring and developing ideas.
\projname{} provides on-demand LLM-based support aimed to inspire the user, rather than guiding the user towards certain directions.
Future systems can further mitigate these risks by improving model explainability by including the confidence level of generated statement and citing original text from the source.
}

\subsubsection{\revise{\textbf{Ethical Implications and Intellectual Ownership}}}
\revise{An ethical concern that arises with human-AI collaborative ideation is the controversy around the intellectual ownership of the work and ideas produced \cite{Smits2022GenAIIP}.
While LLMs can assist researchers in synthesizing information and generating connections between ideas, their role in the creative process raises questions:
if an LLM significantly influences the framing or direction of a research endeavor, to what extent can the resulting ideas be considered original contributions of the researcher? This also poses broader ethical questions about attributing co-authorship to machines: should LLMs be acknowledged as contributors in academic works? While current norms do not recognize machines as authors, there is a need for clearer guidelines on how to appropriately credit the influence of AI systems in research ideation without undermining the originality of human contributions.
In addition, the LLM outputs, fundamentally derived from pre-existing data, might not accurately attribute the presented ideas to the original source document, which could lead to concerns regarding academic integrity and plagiarism.}

\revise{There is also the risk of hindering human creativity due to over-reliance on LLMs \cite{buccinca2021overreliance, Vasconcelos2022ExplanationsCR, kumar2024humancreativityagellms}. 
If researchers begin to lean heavily on machine-generated suggestions, they may become less engaged in the iterative, reflective processes essential for developing innovative and impactful research. 
Such reliance could result in a superficial exploration of research ideas, with researchers focusing only on what is directly surfaced by the LLM while potentially overlooking subtler, more nuanced insights that require human intuition. 
From our user study, participants remained cautious employing LLM-generated ideas and generally examined the idea's validity before incorporating them onto the canvas. 
However, non-CS researchers might be less familiar with the potential pitfalls of relying on LLMs. The system should be explicit about the limitations in its generation.
Ensuring that researchers maintain agency and actively critically evaluate generated outputs is crucial to avoiding this pitfall.}

\revise{To address these concerns, \projname{} does not aim to automate the ideation process or replace the human researcher. Instead, our system prioritizes providing users with controls and presenting relevant information grounded in literature to assist in user-driven ideation.
Future work could further explore strategies for integrating LLM outputs with researcher-led reflection, ensuring that the machine-generated suggestions are contextualized and critically evaluated. 
This would not only safeguard the depth and originality of the research but also preserve the intellectual agency of human researchers, fostering a truly collaborative rather than substitutive dynamic between humans and AI systems.}

\subsection{Limitations \revise{and Future Work}}
\subsubsection{\revise{Generalizability and Domain Customization}}
\revise{Our studies involved participants from specific research areas in HCI and NLP, limiting the generalizability of our system and findings. We also observed that participants from each domains might have different workflows.
For example, some participants from more technical domains, such as algorithm research, preferred writing equations by hand, highlighting the need for future systems to support diverse research methods. While our participants with diverse research topics within HCI and NLP domains benefited from our general approach, conducting further studies to cater to different areas of researchers could examine the system's ecological validity across a broader range of disciplines.}

\revise{One of our key design feature is supporting users in structuring their research ideas using idea nodes. Participants found values in using the structure as a scaffold to further develop their ideas by exploring alternatives for different facets. 
However, while the current predefined structure is generalizable across multiple research domains and scenarios, some participants desired even more structures in the form of finer-grained facets or by defining their own customized facets (H10, N3).
This observation points to interesting future directions of exploring customizable idea facets. Broadly, we envisioned two ways of supporting this. Firstly, we can define a more comprehensive taxonomy of research idea facets, and allow users to dynamically create hierarchies for their ideas. For example, under the current \emph{proposed design and solution}, user could additionally create children nodes for \emph{system architecture}, \emph{interaction design}, etc.  Alternatively, we could also allow users to define their own facets as they ideate in \projname{}. Helping user maintain a useful and manageable structures and provide facet-specific AI supports remains exciting open challenges for future research.}

\subsubsection{Evaluation of Research Ideas}
Estimating the quality of a research idea is challenging due to the uncertain nature of research progression, and even fully executed projects can be hard to predict for continuous impact~\cite{Si2024CanLG,takagi2023towards,kang_augmenting_2022-1, Licuanan2007IdeaEE, Ferioli2010UnderstandingTR}. Personal research interests can also influence how ideas are judged~\cite{gu_generation_2024}. 
Therefore, we \revise{scope this project on the early idea development process and take a user-centered perspective,
focusing on how \projname{} affected researchers' ideation process based on participants' self-perception.}
The short timeframe of the lab study limits our ability to assess the long-term impact of \projname{} on research outcomes. 
\revise{While our longer field deployment uncovered additional strategies participants employed to enhance real-world idea development, assessing the long-term quality, impact, or success of ideas generated through system usage remains a challenge. 
Future work could explore recruiting expert researchers who are experienced in evaluating early research ideas (e.g. senior professors) to assess the early ideation artifacts and predict the idea's impacts.
However, to fully examine the outcome of LLM-assisted research ideation, future works should conduct end-to-end longitudinal studies incorporating public deployments or diary studies, focusing on tracking the lifecycle of ideas from inception to implementation and their eventual outcomes in real-world contexts. For example, Si et al. conducted a large scale study where NLP experts evaluated LLM- and human-generated ideas, and they proposed a follow-up end-to-end study where researchers are recruited to execute the ideas into full projects \cite{Si2024CanLG}. Such efforts would provide deeper insights into the ecological validity of the system and its sustained utility over time.}

\subsubsection{Paper Discovery}
Since research ideas rarely emerge in isolation, participants started ideation with both their initial ideas and any relevant papers they found. While \projname{} used basic tools like search and a recommendation system to expand the paper set, the process remained largely manual. 
\revise{Users were responsible for the paper curation as an attempt to mitigate LLM hallucination on source information and ground generations in real paper contents.}
However, some participants felt the scope of recommended papers was too limited to their existing collection, making it harder to find additional relevant work. This suggests that while focused idea development is important, enabling broader exploration \revise{in the knowledge space} might also be valuable. Future work should balance broader paper discovery with focused, specific feedback to avoid idea drift while supporting further development.

\section{Conclusion}
While many existing research ideation tools focused on supporting broad initial ideation, this paper focused on supporting the iterative expansion and deeper refinement stages of research ideation. 
To bridge this gap, we introduced \projname{}, a research idea development system that provides iterative support for idea refinement and evaluation. \projname{} delivers literature-based feedback grounded by user-collected papers, helping researchers articulate idea facets, including research problems, solutions, evaluations, and contributions. Our system's node-based canvas enables users to visualize variation of idea facets, iteratively refine them, and eventually connect different nodes to compose a coherent research brief.
Our lab study demonstrated that participants using \projname{} explored more idea variations and expanded their initial ideas with greater detail compared to a strong baseline. The deployment study further reinforced the system's utility in more realistic scenarios, with participants reported benefits using \projname{} across various stages of their real-world research projects. These findings underscore the importance of tailoring AI support for different ideation stages to flexibly iterate on and develop research ideas.

\bibliographystyle{ACM-Reference-Format}
\bibliography{references}

\appendix
\newpage
\onecolumn
\lstset{basicstyle=\ttfamily\footnotesize,breaklines=true,breakindent=0pt,breakatwhitespace=true,frame=single}

\section{Demographic Details of Formative Study Participants}
\label{a:formative-participants}
\begin{table}[h]
\small
\centering
    \begin{tabular}[h]{p{1cm} p{1.5cm} p{1.5cm} p{1.5cm} p{1.5cm} p{3cm} p{2cm}}
    \toprule
    P\# & Gender & Age Range & Country & Research YoE & Research Area & AI Use Frequency\\
    \midrule 
    P1 & Man & 25--34 & U.S. & 6--10 & Human-AI interaction & A couple times\\
    P2 & Man & 25--34 & U.S. & 6--10 & LLM evaluation & Occasionally\\
    P3 & Woman & 18--24 & U.S. & 2--5 & LLMs + society & Never\\
    P4 & Man & 25--34 & Canada & 2--5 & Multilingual NLP & Occasionally\\
    P5 & Woman & 25--34 & U.S. & 6--10 & Human-AI interaction & Occasionally\\
    P6 & Man & 25--34 & U.S. & 2--5 & Multimodal AI \& HCI & Occasionally\\
    P7 & Woman & 25--34 & South Korea & 2--5 & LLM retrieval & Occasionally\\
    P8 & Man & 25--34 & U.S. & 6--10 & NLP \& HCI & Occasionally\\
    \bottomrule
    \end{tabular}
    \caption{Participants from our formative study. \textbf{Country} refers to the country in which the participant primarily conducts research at the time of the study. \textbf{Research YoE} refers to the years of experience conducting academic research. \textbf{AI Use Frequency} refers to how frequently the participants uses AI tools to ideate and/or iterate on research ideas. Participants selected ``Occasionally'' based on the description ``I don't rely on AI but sometimes tinker with it.'' }
    \label{t:participants-formative}
\end{table}

\section{Formative Study Wizard-of-Oz Prototype LLM Prompt}
\label{a:formative-prompt}
\begin{lstlisting}
You will be given a brief description of a researcher's research interests and attachments with their most recent publications.
Generate 3 novel research ideas based on the information. The format of each idea should be as follows:

{idea title}
{a super concise and specific bulleted list describing the idea. Include no more than 3 bullet points.}

Here is the brief description of research interests:
{user_research_interest_from_survey}
\end{lstlisting}

\section{Formative Study Qualitative Analysis Themes}
\label{a:codebook}
\begin{table}[h]
\small
\centering
    \begin{tabular}[h]{p{3cm} p{11cm}}
    \toprule
    Theme & Description \\
    \midrule 
    Ideation process & How participants formulated initial ideas, reviewed and synthesized literature, backtracked and changed direction of their ideas, evaluated and converged on an idea, and developed their own approaches to ideation. \\
    Workflow challenges & Challenges encountered by participants during ideation, related to areas such as managing ideas and information, evaluating ideas, scoping ideas, effectively communicating/gathering feedback on ideas, and reviewing literature. \\
    Artifacts generated & Characteristics of artifacts participants generated throughout their ideation and research process, both in their own work outside of the study as well as during the study. \\
    Use of AI & How participants used AI and envisioned to use AI, both in their own work outside of the study as well as during the study. Limitations and avoidance of AI were also included. \\
    Feedback & Participants' feedback on the interactive capabilities of the probe and responses provided by the AI system, as well as new features and interactions they wanted to see going forward.\\
    
    \bottomrule
    \end{tabular}
    \caption{Themes derived from a qualitative analysis of our formative study transcripts and recordings.}
    \label{t:participants-formative}
\end{table}

\section{Demographic Details of Comparative Lab Study Participants}
\label{a:lab-participants}
\begin{table}[H]
\small
\centering
    \begin{tabular}[h]{p{1cm} p{1.5cm} p{1.5cm} p{1cm} p{1.5cm} p{3.5cm} p{2cm}}
    \toprule
    P\# & Gender & Age Range & Country & Research YoE & Research Area & AI Use Frequency\\
    \midrule 
    H1 & Woman & 25--34 & Canada & 6--10 & HCI + AI & Occasionally\\
    H2 & Man & 25--34 & Canada & 6--10 & Mixed Reality & Never\\
    H3 & Woman & 25--34 & U.S. & 2--5 & HCI + AI & Occasionally\\
    H4 & Man & 18--24 & U.S. & 2--5 & Fabrication & A couple times\\
    H5 & Man & 18--24 & U.S. & 6--10 & Ubiquitous Computing & Always\\
    H6 & Man & 25--34 & Canada & 2--5 & Interaction Techniques & A couple times\\
    H7 & Woman & 25--34 & U.S. & 6--10 & HCI + Health & A couple times\\
    H8 & Man & 25--34 & U.S. & 2--5 & HCI + Generative AI & Always\\
    H9 & Man & 25--34 & U.S. & 2--5 & HCI + AI & Always\\
    H10 & Man & 25--34 & U.S. & 6--10 & Computational Social Science & A couple times\\
    H11 & Non-binary & 25--34 & U.S. & 2--5 & HCI + Culture & Never\\
    H12 & Woman & 25--34 & Canada & 6--10 & Haptics + Robotics & Frequently\\
    \midrule
     N1 & Woman & 18--24 & U.S. & 2--5 & Language Modeling & Frequently\\
    N2 & Man & 25--34 & Canada & 6--10 & LLM Interpretability & Always\\
    N3 & Man & 25--34 & U.S. & 2--5 & Conversational Agents & Always\\
    N4 & Man & 18--24 & U.S. & 2--5 & NLP Fairness + Law  & A couple times\\
    N5 & Woman & 25--34 & U.S. & 2--5 & LLM Alignment + Security & Never\\
    N6 & Man & 25--34 & U.S. & 2--5 & Multimodal LLMs & A couple times\\
    N7 & Woman & 25--34 & U.S. & 2--5 & LLM Interpretability & Never\\
    N8 & Woman & 18--24 & U.S. & 0--1 & Educational Data Science & Frequently\\
    \bottomrule
    \end{tabular}
    \caption{Participants from our laboratory study. \textbf{P\#} is the participant ID. \textbf{Country} refers to the country in which the participant primarily conducts research at the time of the study. \textbf{Research YoE} refers to the years of experience conducting academic research. \textbf{AI Use Frequency} refers to how frequently the participants uses AI tools to ideate and/or iterate on research ideas. Participants selected ``Occasionally'' based on the description ``I don't rely on AI but sometimes tinker with it.'' }
    \label{t:participants-lab}
\end{table}

\section{Comparative Lab Study User Task Instruction}
\label{a:task-instruction}
\begin{lstlisting}
User Task Sheet
Welcome to the Research Ideation Session!
You will be working on research tasks within the domain of [Task Theme]. Below, you'll find the details of your assigned tasks, along with important criteria to keep in mind as you develop your ideas.

Task Overview
You will be assigned one of the following broad topics:
Task 1: [Task Topic 1]
Task 2: [Task Topic 2]

Your Goal:

You will have 30 minutes to complete each task. The focus of the study is on the research ideation process.
For each task, explore research ideas and literature to draft a research brief of around 12 sentences. 

Your brief should include the following sections:
Problem Description and RQ
Proposed Design and Solution
Evaluation Method
Contribution and Impact

Evaluation Criteria:
As you work on your tasks, consider the following criteria for your research brief:

Novelty
Definition: The research brief justifies the research idea's originality by clearly differentiating it from prior works.

Impact
Definition: The research brief describes ideas that make a nontrivial contribution to advance knowledge in the field.

Specificity
Definition: The research brief is clearly elaborated and sufficiently detailed with a complete description of the research idea(s).

Feasibility
Definition: The research brief describes the implementation of the ideas in a reasonable way that does not violate known constraints.

Relevance
Definition: The research brief addresses the task topic at hand and can be reasonably expected to apply to the research problem.

Please ensure that your research brief reflects these attributes.

Please click on the System URL and the User Survey to start.

Good luck, and enjoy the ideation process!
\end{lstlisting}

\section{Comparative Lab Study Qualitative Analysis Themes}
\label{a:codebook-lab}
\begin{table}[H]
\small
\centering
    \begin{tabular}[h]{p{3cm} p{11cm}}
    \toprule
    Theme & Description \\
    \midrule 
    Comparison of system and workflows & How participants compared our \projname{} system with the baseline system, as well as comparisons of their workflow in \projname{} with their existing or past workflows without \projname{}.\\
    Use cases & Comments on the specific use cases of \projname{}. Examples include idea generation, literature review, sharing research ideas, creating mind maps, onboarding new researchers, and more. \\
    General system benefits & Participants' perceived benefits of \projname{}. Examples include improving literature understanding, more thoroughly expanding idea facets, refining ideas through iteration, exploring ideas through alternatives, preventing idea fixation, and more. \\
    Usage of node-based canvas interface & How participants made use of \projname{}'s node-based canvas interface. Includes comments about the use of individual nodes, node edges, selecting a group of nodes and generating a research brief, and AI suggestions stemming from nodes. \\
    System limitations & Participants' comments about \projname{}'s limitations and drawbacks, most notably drawbacks of the node-based canvas interface, broadness of LLM suggestions, misalignment between LLM output and participant's idea, and lack of trust towards AI suggestions/summaries. \\
    Future directions & Participants' suggestions for directions to improve \projname{} as well as addressing tradeoffs and ethical concerns of using the system.\\
    
    \bottomrule
    \end{tabular}
    \caption{Themes derived from a qualitative analysis of our lab study transcripts and recordings.}
    \label{t:participants-formative}
\end{table}

\section{Comparative Lab Study Survey Response}
\label{a:survey-response}
\begin{longtable}{|p{11cm}|p{4cm}|}
\hline
\multicolumn{2}{|c|}{\makecell{\textbf{On a scale of 1-Strongly disagree, 2-Disagree, 3-Slightly disagree, 4-Neutral, } \\ \textbf{5-Slightly agree, 6-Agree, 7-Strongly agree, how much do you agree with the following statements:}}} \\
\hline
The research brief justifies the research idea's originality by clearly differentiating from prior works. & Novelty \\
\hline
The research brief describes ideas that make a nontrivial contribution to advance the knowledge in the field. & Impact \\
\hline
The research brief is clearly elaborated and sufficiently detailed with the complete description of the research idea(s). & Specificity \\
\hline
The research brief describes the implementation of the ideas in a reasonable way that does not violate known constraints. & Feasibility \\
\hline
The research brief addresses the task topic at hand and can be reasonably expected to apply to the research problem. & Relevance \\
\hline
I felt confident in my understanding of the existing literature related to my research idea. & Confidence in Literature \\
\hline
I was able to effectively expand on my initial research idea with more details. & Expanding Research Idea \\
\hline
I was able to sufficiently explore and evaluate alternative versions of my research ideas. & Exploring Research Ideas \\
\hline
I was able to create a well-structured and coherent research brief. & Creating Research Brief \\
\hline
The system provides suggestions that are specific to my research ideas and the related literature. & Specific Suggestions \\
\hline
The system provides suggestions that are actionable for me to advance my research ideation process. & Actionable Suggestions \\
\hline
How mentally demanding was the task? 1-Very low, 7 Very high & Mental Demand \\
\hline
How physically demanding was the task? 1-Very low, 7 Very high& Physical Demand \\
\hline
How hurried or rushed was the pace of the task? 1-Very low, 7 Very high& Pace of Task \\
\hline
How successful were you in accomplishing what you were asked to do? 1-Very low, 7 Very high& Success in Accomplishing Task \\
\hline
How hard did you have to work to accomplish your level of performance? 1-Very low, 7 Very high& Work to Accomplish Performance \\
\hline
How insecure, discouraged, irritated, stressed, and annoyed were you? 1-Very low, 7 Very high& Insecure, Discouraged, Irritated, Stressed, and Annoyed \\
\hline
I felt that I was in control when I used the system. & Control When Using System \\
\hline
I felt like the system allowed me to make choices and decisions while developing the research idea. & Choices and Decisions \\
\hline
I trust the system's generated research ideas to be used in a real research scenario. & Trust in Research Ideas \\
\hline
I trust the system's AI response to my prompt. & Trust in AI Response \\
\hline
I trust the system's reference and summary for the related literature. & Trust in Literature \\
\hline

\caption{Survey questions and their corresponding labels}
\label{tab:survey_labels}
\end{longtable}

\begin{figure}[H]
    \centering
    \includegraphics[width=\textwidth,keepaspectratio]{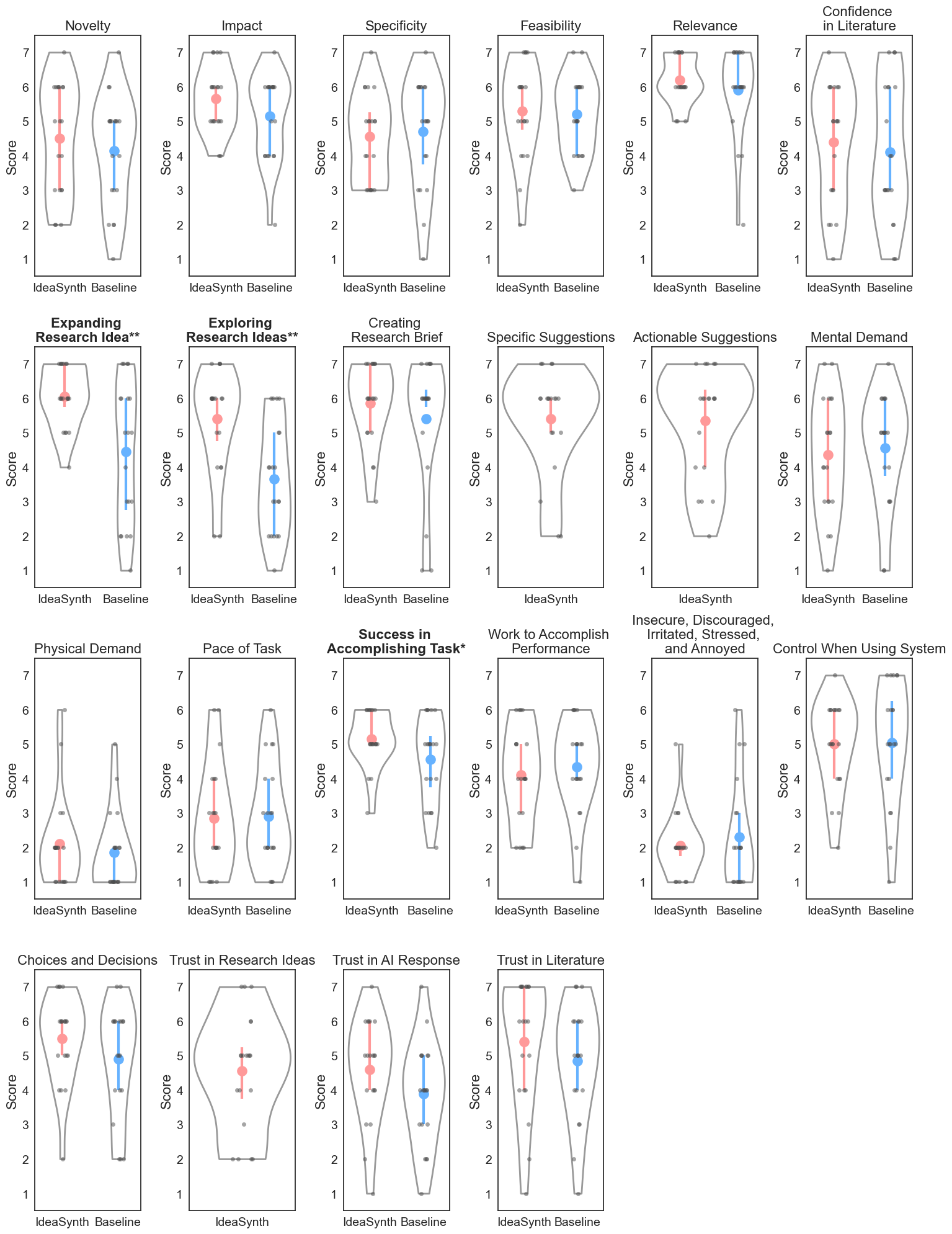}
    \caption{
        \textbf{Survey Response.}
        \textnormal{Bolded plots show statistically significant differences ($p<0.05^{*}, p<0.01^{**}$) between \projname{} and baseline, using Wilcoxon signed-rank test if each survey question is treated independently. However, to avoid multiple comparison bias, we only report pre-selected measures (Fig.\ref{fig:survey-MANOVA}) before analyzing the survey results.}
    }
    \label{fig:survey-res}
\end{figure}

\section{Deployment Study Qualitative Analysis Themes}
\label{a:deployment-codebook}
\begin{table}[h]
\small
\centering
    \begin{tabular}[h]{p{3cm} p{11cm}}
    \toprule
    Theme & Description \\
    \midrule 
    Task and Tool Usage & How participants utilized \projname{} in their real-life projects, the stages of ideation they used the tool for, and the outcome of the tool usage. \\
    Helpful Features & Significant system functionalities that assisted participants, and the specific benefits resulted from those features.\\
    Future Use Case and Workflow & Envisioned use case of \projname{} longitudinally, how the tool can be applied in real-life research setting, and how the tool can be fit into participants' workflows. \\
    Concerns with LLM in Research & Trust towards LLM generated literature summary, concern about evaluating LLM-generated research ideas, and tension on user control. \\
    System Suggestions & Expanding literature scope to suggest more papers, customizability to define idea facets based on the research domain, and adaptive assistance for different parts of the research process. \\
    
    \bottomrule
    \end{tabular}
    \caption{Themes derived from a qualitative analysis of our deployment study transcripts and recordings.}
    \label{t:theme-deployment}
\end{table}

\section{\projname{} LLM Prompts}
\label{section:llm-prompts}

\subsection{System Prompt}
\begin{lstlisting}
You are a helpful research assistant. Your task is to assist a researcher in developing their initial research idea into a comprehensive research brief during a research ideation session.
The research brief should include the following facets:
'Problem Description and RQ', 'Proposed Design and Solution', 'Evaluation Method', 'Contribution and Impact'.
These facets are represented as nodes, and are interconnected, with each connection representing the logical relationship between them. 
The entire path of research facets and their connections will be used to draft the final research brief.
\end{lstlisting}

\subsection{Paper Processing Prompt}
\label{prompt:paper-processing}
\begin{lstlisting}
Answer a question based on the following excerpts from the full text of the paper.

{full_text}

Given the information above, please summarize the following aspects of the paper each in a paragraph: 
Problem Description and RQ, Proposed Design and Solution, Evaluation Method, Contribution and Impact, Limitation and Future Work. 
Respond in proper JSON format like this {"key": "value"}.
If the question cannot be answered from the provided context, reply {"No answer": ""}
\end{lstlisting}

\subsection{Node Suggestion Prompt}
\label{prompt:node-suggestion}
\begin{lstlisting}
Idea Facet: {idea_facet}

Research Idea: 
Title: {node_title}
Content: {node_content}

Here are related literatures, ONLY use these provided papers:
{papers_data}

Your main tasks are:

1. Regenerate Current Idea Facet: Ask clarifying questions to help users reflect on their description. Provide evidence to support or challenge the user's claims.

2. Generate Alternatives: If the user wishes to expand on the current facet, generate a few alternative versions based on their feedback or selected context.

3. Generate New Idea Facets: Help generate new facets that are specific and relate to the existing facets and idea context when the user wants to expand their idea.

Provide your response in the following format:
{{
    "ai_suggestion": [
        {{
            "idea_facet": [Idea Facet to be act on],
            "action": [Action that the user can take or the system should perform, from the list above, based on the input data],
            "suggestion": [Proposed Suggestion, whether it is questions for the user, supporting information, or suggested edits, if action is targeted to a specific idea facet, the suggestion should name that facet. If the suggestion references a paper, add a citation tag in the format of @ref[corpusId].],
        }},
        ...
    ]
}}

 Here are two example outputs:
 {{
    "ai_suggestion": [
        {{
            "idea_facet": "Problem Description and Research Question",
            "action": "Regenerate Current Idea Facet",
            "suggestion": "Could you elaborate on the specific features of your system that are hypothesized to enhance [the task context]? How do these features align with the findings from studies like 'Paper2'@ref[corpusId] and 'Paper1'@ref[corpusId]?""
        }},
        {{
            "idea_facet": "Proposed Design/Solution",
            "action": "Generate New Idea Facets",
            "suggestion": "Maybe we can start to think about the next facet, Evaluation Method. How can we design the metrics and experimental setup to measure the effectiveness of the proposed system's features?"
        }}
    ]
}}
    
You may include multiple suggestions, like in the example, if necessary, but each suggestion should only be 1-2 sentences and easy to read. Always respond in a clear and concise manner, focusing on the specific idea facet or task at hand. Do not summarize or reiterate what the user has said.
Begin your response by identifying which idea facet the user is focusing on, and then proceed with the appropriate task based on their input. Reference the related literature to provide evidence or context for your suggestions.
\end{lstlisting}

\subsection{Node Generation Prompt}
\label{prompt:node-generation}
\begin{lstlisting}
You will be generating a new research idea facet based on the user's existing idea facet and context. Your goal is to create a specific and relevant facets that contributes to the overall research brief.
Here is the current idea facet:
{current_node_data}

And here is the context for the new idea facet:
{node_context}

Generate new idea facet(s) based on the following steps:
1. Review the user's current idea facet and the context of all connected nodes.
2. Identify key themes, objectives, or gaps that can be addressed by the new facet.
3. Generate a new idea facet that is specific, relevant, and logically connected to the current idea facet.
4. Ensure that the new facet contributes to the overall coherence and completeness of the research brief.

Here are the instructions:
{action_to_take}

Use the suggestion to guide the generation of the new idea facet(s).
If the instruction for Action is Regenerate Current Idea Facet or Regenerate Current Node, use the suggestion from existing facet and the user's feedback to generate ONLY 1 refined version of the existing facet.
If the instruction for Action is Generate Alternatives, generate up to 3 alternatives of the same type as the original facet to ensure that the new facets offers distinct perspectives or focuses. 
If the instruction for Action is one of 'Problem Description and RQ' or 'Proposed Design and Solution' or 'Evaluation Method' or 'Contribution and Impact', generate up to 3 new nodes of ONLY that specific facet type and clearly establish the logical connection to the original facet.

Provide your response in the following format:
{{
    "new_nodes": [
        {{
            "ideaFacet": [New Idea Facet Type],
            "title": [Title of the new idea facet in less than 10 words],
            "content": [Content of the new idea facet in 1-3 short sentences, use the (author, year) format to refer to paper without the corpusId],
        }},
        ...
    ]
}}

ideaFacet should be one of the following: 'Problem Description and RQ', 'Proposed Design and Solution', 'Evaluation Method', 'Contribution and Impact'.
Ensure that each new idea facet is clearly defined and contributes to the overall research brief. Your response should be concise, focused, and directly relevant to the user's input.
\end{lstlisting}

\subsection{Node Edge Generation Prompt}
\label{prompt:edge-generation}
\begin{lstlisting}
You will be establishing connections between two different idea facets provided by the user. Your goal is to create logical relationships between the facets that form a coherent research brief.

Here are the two idea facets you need to connect:
{source_node_data}

{target_node_data}

To establish a connection between the two facets, follow these steps:
Analyze facets:
- Identify main ideas and key elements.
- Determine strengths and limitations.
Identify Connections:
- Find common themes or complementary aspects.
- Consider how one facet could enhance or support the other in the research process.
Suggest Enhancements:
- Provide specific ways to refine the connection, with examples or references.

Format your response as follows, replacing the placeholders with the relevant information:
{{
    "connectionStrength": [a number value from 0 to 1 indicating the coherence of the connection, 1 being the strongest],
    "suggestion": [Start with a statement, such as "The connection is weak/average/strong". Then use one sentence to offer suggestion for enhancing the connection. Make your point concise and clear],
}}

Here are two example responses:
{{
    "connectionStrength": 0.8,
    "suggestion": "The connection is relatively strong. Consider making the measureable outcomes, such as task completion time, explicit in the evaluation methods to directly link back to metrics that can be extracted from the proposed design."
}}
{{
    "connectionStrength": 0.3,
    "suggestion": "The connection is weak. The dataset in your proposed method might not offer the user interaction granularity that is required to answer the research questions. Consider framing the questionis more broadly, like in paper @ref[corpusId]."
}}

Your response should be concise, clear, and focused on the logical relationship between the two facets. Aim to provide actionable suggestions that help the user improve the coherence and flow of their research brief.
connectionStrength should be a number between 0 to 1 inclusive, be fair and critical about the coherence between the facets connected, if the facets are not specific enough to evaluate the connection, rate it with a low number and suggest further clarifcation. Be strict and use the whole range of values so the differences between strong and weak connections are distinct. Do not use 0.5 as that is the default neutral value.
If papers are referenced, use the citation tag in the format of @ref[corpusId].
\end{lstlisting}

\subsection{Research Brief Generation Prompt}
\label{prompt:brief-generation}
\begin{lstlisting}
You will create a one page academic research brief based on the provided idea facet nodes, their connecting edges, and literature references to collected papers. Your goal is to create a coherent and structured research brief that integrates the facets logically and effectively.
Here are the idea facet nodes and their connections:
{research_ideas}

Here are the collected papers for literature references, ONLY use these provided papers:
{papers_data}

Follow these steps to generate the research brief:
1. Review the idea facet nodes and their connections to understand the logical flow of the research brief.
2. Identify literature references from the collected papers that support or relate to each idea facet.
3. Develop a one page academic research brief that includes the following sections in this order, each section should be around 3 sentences:
    - Title: A concise and descriptive title for the research brief.
    - Problem Description and Research Question: Connect and expand on the content covered in Problem Description and RQ nodes.
    - Proposed Design and Solution: Connect and expand on the content covered in Proposed Design and Solution nodes.
    - Evaluation Method: Connect and expand on the content covered in Evaluation Method nodes.
    - Contribution and Impact: Connect and expand on the content covered in Contribution and Impact nodes.

Include in-text citations to the literature references (e.g. [1, 3, 5] refer to the first, third and fifth publication in the literatureReferences) where relevant throughout different sections. Ensure that the research brief is coherent, structured, and effectively integrates the idea facets with the literature.

Keep as much of the text in the node title and content as possible. Only make minor additions or modifications to connect and expand on existing nodes.

Provide your response strictly in the following format:
{{
    "researchBrief": {{
        "title": [Title of the research brief],
        "problemDescription": [Problem description and research question section of the research brief],
        "proposedDesign": [Proposed design and solution section of the research brief],
        "evaluationMethod": [Evaluation method section of the research brief],
        "contributionImpact": [Contribution and impact section of the research brief],
    }},
    "literatureReferences": [
        {{
            "citation_id": [id of the citation used to refer to the paper in the research brief],
            "paper_title": [Full Paper Title],
        }},
        ...
    ]
}}

Ensure that the research brief is coherent, structured, and effectively integrates the idea facets with relevant literature references. 
Only respond with the JSON object as specified above, do not include any additional information or summaries.
\end{lstlisting}

\subsection{AI Chat Q\&A Response Prompt}
\label{prompt:ai-chat-qa-response}
\begin{lstlisting}
You will be generating a response based on the user's prompt, the collected paper information and summary, and the user's research ideas. Your goal is to provide a concise and informative response that addresses the user's prompt and references the relevant literature.
Here is the user's prompt:
{user_prompt}

Here are the papers data, ONLY use these provided papers:
{papers_data}

Here is the user's research idea:
{research_ideas}

Based on the prompt, the papers, and the research ideas, generate a concise response in the following format:
{{
    "litReviewResponse": "Response to the user's prompt if provided. If the response references a paper, add a citation tag in the format of @ref[corpusId]"
}}
Only respond with the JSON object as specified above, do not include any additional information or summaries.
\end{lstlisting}

\subsection{Literature Review Summary Prompt}
\label{prompt:lit-review-summary}
\begin{lstlisting}
You will be generating a concise one-paragraph literature review based on provided list of corpus_id:
{corpusIds}

Here are the papers data, ONLY use these provided papers:
{papers_data}

Refer to the papers by their short-form titles with citatioin format (AuthorLastName et al., year) and focus on the key method, findings, limitations, and future works of the papers. 
Output in the following format:
{{
    "litReviewSummary": "One paragraph summary of the literature review based on the provided papers and prompt. If the summary references a paper, add a citation tag in the format of @ref[corpusId]"
    "corpusIds": [List of corpusIds of the mentioned papers to the user's prompt.]
}}
Only respond with the JSON object as specified above, do not include any additional information or summaries.
\end{lstlisting}

\subsection{Literature Review Analysis Prompt}
\label{prompt:lit-review-anlaysis}
\begin{lstlisting}
You will be generating a literature analysis based on provided paper summaries and current research ideas.
First, here are the summaries of relevant papers, ONLY use these provided papers:
{paper_summaries}

Next, here are the current research ideas:
{research_ideas}

Analyze the paper summaries and research ideas:
1. Identify the main themes and concepts present in both the papers and research ideas.
2. Look for connections, similarities, and differences between the existing literature and the proposed research ideas.
3. Consider how the existing literature supports or challenges the research ideas.

Generate a concise literature review with the following 2 sections:
1. Highlight key sections from the papers that are most relevant to the research ideas, whether it's research problem statement, method, evaluation, or contribution and impact. Discuss how the section relates to the proposed research ideas, noting any knoweldge gaps in existing works or areas where the new ideas extend current knowledge.
2. Conclude next steps for the researcher to address based on the literature review and the proposed research ideas, provide actionable steps like pivot the research question, refine existing methods, or develop evaluation.

Format your response as follows, remove the placeholders and replace them with the relevant information:
{{
    "analysis": [
        {{
            "section_title": "[One-line summary of the relevance between the paper section and the research idea.]",
            "paper_title": "[Full Paper Title. Add a citation tag in the format of @ref[corpusId]]",
            "corpus_id": "[Corpus ID 1]",
            "key_section": "[3-5 sentences description of the relevant content in the key section, don't just describe a framework/method/result was proposed, but what it is about.]",
            "connection_to_ideas": "[2-3 sentences description of how this section relates to the proposed research ideas, name the specific idea being discussed.]",
            "next_steps": [
                "[Actionable steps for the researcher to address based on the literature review and the proposed research ideas.]",
                ...
            ],
        }},
        ...
    ],
}}

Only output the above JSON object. Ensure that your literature analysis is concise, coherent, and effectively synthesizes the information from both the paper summaries and research ideas. Refer to the papers by their short-form titles with citatioin format (AuthorLastName et al., year) in key_section and connection_to_ideas.
\end{lstlisting}

\subsection{Node Literature Review Prompt}
\label{prompt:node-lit-review}
\begin{lstlisting}
You will be analyzing the user's research idea based on related literature. Your goal is to provide a comprehensive literature-based analysis of the particular research idea, identify the most relevant papers and sections, and suggest refinements based on the existing literature.

Here's the user's initial research idea:
Idea Facet: {idea_facet}
Title: {title}
Content: {content}

Now, review the following list of related papers, including their titles, TLDRs, abstracts, and section summaries, ONLY use these provided papers:
{paper_summaries}

Consider the following summary of all related works in relevance to the user's ideas:
{lit_review_summary}

To analyze the research idea based on the literature, follow these steps:
1. Reading the user's research idea, related papers, and summaries.
2. Identifying key concepts and themes in the idea and comparing these with the related papers.
3. Determining the most relevant papers based on research questions, methodologies, and findings.
4. Highlighting specific sections in these papers that directly relate to the idea.
5. Suggesting refinements to the research idea by addressing gaps, incorporating methods, adjusting scope, or considering alternative approaches.

Your output has three facets:
most_relevant_sections: List the 1-3 most relevant sections from papers, explaining why each is particularly important to the research idea. Include the paper title and a brief explanation of its relevance. For each of the most relevant papers, identify the key section that are particularly important to the research idea. Explain why these sections are significant and how they relate to the proposed research.
suggestions: Offer 1-2 specific suggestions for refining or improving the research idea based on your analysis of the literature. Explain the rationale behind each suggestion and how it could strengthen the research.

Remember to support your analysis with specific references to the papers and their contents. Be objective in your assessment and provide constructive suggestions for improvement.

Present your analysis in the following format in JSON:
{{
    "most_relevant_sections": [
        {{
            "section_title": "[One-line summary of the relevance between the paper section and the research idea.]",
            "paper_title": "[Full Paper Title, add a citation tag in the format of @ref[corpusId]]",
            "key_section": "[3-5 sentences description of the relevant content in the key section, don't just describe a framework/method/result was proposed, but what it is about.]",
            "connection_to_ideas": "[2-3 sentences description of how this section relates to the proposed research ideas, name the specific idea being discussed.]"
        }},
        ...
    ],
    "suggestions": [
        "[Actionable steps in 2-3 sentences for the researcher to address based on the literature review and the proposed research ideas.]",
        ...
    ]
}}

Only output the above JSON object and NOTHING ELSE. Ensure that your literature analysis is concise, coherent, and effectively synthesizes the information from both the paper summaries and research ideas. Refer to the papers by their short-form titles with citatioin format (AuthorLastName et al., year) in key_section and connection_to_ideas.
\end{lstlisting}

\end{document}